\documentclass[preprint, amsmath]{revtex4}
\usepackage{amssymb}
\usepackage{indentfirst}
\usepackage[mathscr]{eucal}
\usepackage{graphicx}
\usepackage[T2A]{fontenc} 
\usepackage[cp1251]{inputenc} 
\usepackage{srcltx} 

\begin{document}

\title{
 Polarons on one-dimensional lattice. \\
        II. Moving polaron.
      }

\author{
          T.Yu. Astakhova, V.N. Likhachev, G.A. Vinogradov${}^*$
       }

\affiliation{
 Emanuel Institute of Biochemical Physics,
 Russian Academy of Sciences,
 ul. Kosygina 4,  Moscow 119334, Russia\\
${}^*${\rm E-mail: gvin@deom.chph.ras.ru}
            }

\begin{abstract}

In the present study we revise the possible polaron contribution
to the charge and energy transfer over long distances in
biomolecules like DNA. The harmonic and the simple inharmonic
($U(x) = x^2/2 - \beta x^3/3$) lattices are considered. The
systems of PDEs are derived in the continuum approximation. The
PDEs have the one-soliton solution for polarons on the harmonic
lattice. It describes a moving polaron, the polaron velocity lies
in the region from zero to the sound velocity and depends on the
polaron amplitude. The PDEs describing polarons on the inharmonic
lattice also have the one-soliton solution only in the case of
special relation between parameters (parameter of inharmonicity
$\beta$ and parameter of electron-phonon interaction $\alpha$).
Polaron dynamics is numerically investigated in the wide range of
parameters, where the analytical solutions are not available.
Supersonic polarons are observed  on inharmonic lattice with high
inharmonicity. There is the range of parameters $\alpha$ and
$\beta$ where exists a family of unusual stable moving polarons
with the envelope consisting of several peaks (polarobreather
solution). The results are in qualitative agreement with recent
experiments on the charge transport in DNA.

\vspace{3 cm}



\end{abstract}

\maketitle



\section{Introduction}
  \label{sec:Intro}

This paper continues our investigations on the polarons in 1D
lattices \cite{Lik12}. Its primary goal is an explanation of the
high efficient charge transfer in biological macromolecules.

Charge transfer (CT) is of utmost importance in physics, chemistry
and biology \cite{Bar96, Bix99, Ber04, Mat02, Ada03, Wei05, Gen10,
Mal10, Shi10,Kuz95, Kuz99, Bal01}. For instance, it is essential
in processes at nano-- and micro--scales, for solar cells
technology \cite{Was92}, molecular electronics \cite{Rat97, Lin07}
and other novel technologies. The CT is very important, {\it e.g.}
for oxidative DNA degradation \cite{Sch04}. Noteworthy is that DNA
has been labelled as electric conductor \cite{Fin99},
semiconductor \cite{Por00}, insulator \cite{Ber97} and even
superconductor \cite{Kas01}. Such wide range of possible electric
properties is explained by the variety of different DNA structures
used, diversity of experimental methods and inner and outer
surrounding DNA conditions.  It has been shown that the rate of CT
exponentially decays with the distance $L$ as $k_{\rm CT} \propto
\exp(- \gamma L) $, where $\gamma \gtrsim 0.1-0.2$ \AA${}^{-1}$
\cite{Tak04, Lew06a, Lew06b}.

Significant progress in enhancing the conducting properties of DNA
was achieved after the DNA synthesis reached its present status of
automation. Apparently adenine (A) is the most preferable building
block because of its resistance to charge trapping \cite{Lew01}
and its CT efficiency \cite{Nei04, Sha05}. Three sets of duplexes
were synthesized  giving 4--, 6-- and 14--base A tracts with a
covalently attached rhodium complex
[Rh(phi)${}_2$(bpy$'$)]${}^{3+}$ serving as the photooxidant
\cite{Aug07}. It was found that there was no significant change in
degree of decomposition ($\gamma = 0.0013$ \AA${}^{-1}$), i.e. the
efficiency of charge transport very slowly depends on the
distance, what contrasts with the larger values found in previous
experiments. The possibility of CT over 34 nm (100-mer DNA) was
recently announced by J.~Barton and collaborators \cite{Sli11} and
this result surpasses most previous achievements with molecular
wires.

Analogous results were obtained for the synthetic $\alpha$-helical
peptides where a very shallow distance dependence of the charge
transfer was also found \cite{Ari10}. One interesting feature in
these studies was revealed: the charge transport is a coherent
single-step process, i.e. the charge carrier can be transferred
ballistically \cite{Gen11, Bar11}.

Traditionally the hopping and tunnelling are considered as a most
abundant mechanisms in the theory of charge transfer in DNA
\cite{Wag05, Sch04a, Sch04b, Cha07}. Polarons were also
extensively studied as candidates for charge carriers \cite{Con01,
Con00, Con05, Kuc10, Mah10, Wu-11, Feh11, Man03} (see also review
\cite{Dev09}). At present, the polaron theory of CT in
biomacromolecules is very popular. Polarons are considered in
different models and approximations \cite{Kuc10, Kal98, Sch10,
Kub10,  Lak10, Lak05, Hen04, Hen06, Yam07}.

Effects of nonlinearity can give contribution to the charge
transport due to formation of the bounded soliton--electron state
\cite{Zol93, Zol96} or polarobreathers \cite{Aub97, Cru00, Cue06,
Fla98, Hen00, Hen03, Kal04, Kal05, Yu-04}. As a working definition
breathers are localized solutions with at least two degrees of
freedom in a one-dimensional lattice (i.e., they have an
``internal'' degree of freedom, a vibrational motion) while stable
localized solutions with one degree of freedom are solitons
\cite{Cru00}.

Recently the concept of nonlinearity was additionally advanced and
solectrons, -- bounded state of solitons and charge carriers, were
found in numerical experiments (see \cite{Vel10} and references
therein).

Notably, the charge transfer in DNA appearers to be independent of
distance but is critically sensitive to different kinds of
defects, -- static and dynamical disorder \cite{Boo00, Zha04}.
Biomacromolecules are very complex systems and many kinds of
fluctuations can disturb the path-ways of charge migration. A DNA
macromolecule is supposed to have a uniform or a random
distribution of the helix angles. And an influence of angles
fluctuations on the charge transfer efficiency was analyzed
\cite{Qua08, Cra05, Yu-01, Bru00}. Dynamical fluctuations
(temperature) also can affect the hopping probability thus giving
rise to charge localization \cite{Con00, Kat02, Qu-10, Qua09}.

In the present paper we consider the charge transport in harmonic
and inharmonic lattices and derive analytical expressions for
moving polarons in the continuum approximation. Polaron dynamics
is investigated numerically in the range of parameters where the
analytical solutions are not available. Special attention is given
to possible explanation of recent experiments on the charge
transport in DNA.


\section{Moving polaron on the harmonic lattice}
  \label{Harm_Latt}

We start from consideration of 1D harmonic lattice with free
boundaries consisting of $N$ particles with the hamiltonian
\begin{equation}
  \label{1}
  H = \dfrac{m}{2} \sum\limits_{j=1}^{N} \dot x_j^2 +
      \dfrac{k}{2} \sum\limits_{j=1}^{N-1} (x_{j+1}-x_j)^2 +
      \left< \vec\Psi \,| \widehat H^{\rm e} |\, \vec\Psi \right>,
\end{equation}
where $\vec\Psi = \psi_1, \psi_2, \ldots, \psi_N$ is the discrete
wave function,  $m$ is the particle mass, $k$ is the lattice
rigidity, $x_j$ is  deviation of the $j$th particle from the
equilibrium. ``Particle'' represents the DNA base, and their
interaction is due to the  $\pi-\pi$ overlapping of neighboring
heteroaromatic bases.

The components of the electron-phonon interaction operator are:
\begin{equation}
  \label{b1}
 H_{i,j}^{\rm e} = \delta_{j,j} e_j + \delta_{j,j-1} t_{j-1} +
                              \delta_{j,j+1} t_{j},
\end{equation}
where  $\delta_{i,j}$ is the Kronecker symbol;  $t_j$ -- hopping
integral between $j$th and $(j+1)$th particles; $e_j$ is the
energy of interaction of the charge carrier with the lattice site
(on-site energy). This strategy is known as the tight-binding
approach to DNA which takes into account the on-site energies and
suitably parametrized hopping onto the neighboring sites
\cite{Con00, Con05, Cun07}.

The main object of the present investigation is  synthetic DNA
comprised of identical base pairs. So the diagonal disorder is
absent and $e_i =$ const. This constant is the electronic energy
origin and without the loss of generality can be put to zero. The
hopping integral is often written in the form suggested by Su,
Schrieffer and Heeger \cite{Su-79,Su-80}:
\begin{equation}
  \label{c}
  t_j = - [v_0 - \alpha (x_{j+1} - x_j)] \,,
\end{equation}
where  $v_0$ is the hopping integral at the equilibrium and
parameter   $\alpha$ accounts the electron-phonon interaction. The
electron charge transfer is considered, though all results are
valid for the hall transfer.

It is convenient to make hamiltonian \eqref{1} dimensionless using
mass $m$, energy $v_0$ and the rigidity coefficient $k$ as the
units. Then the unit of time  $[t] = \sqrt{m/k}$, unit of length
$[L] = \sqrt{v_0/k}$. If the DNA parameters are chosen
\cite{Con00, Con05}, namely  $v_0 = 0.3$~eV, $k =
0.85$~eV/\AA${}^2$, $m = 130$~a.m.u., then the time unit is $(t)
\approx 0.13$~ps, and the length unit is $[L] \approx 0.59$~\AA.
Parameter $\alpha$ is also dimensionless:  $\alpha/\sqrt{v_0 k}
\to \alpha$. The dimensionless value of this parameter for DNA
$\alpha \approx 1.2$. Parameter $\alpha$ is the single parameter
specifying the lattice.

The dimensionless hamiltonian \eqref{1} reads:
\begin{equation}
  \label{d}
  H = \dfrac12 \sum\limits_{j=1}^N \dot x_j +
      \dfrac12 \sum\limits_{j=1}^{N-1} q_j^2 -
      \sum\limits_{j=1}^{N-1}(1 - \alpha q_j) \,
      (\psi_j^* \psi_{j+1} + {\rm c.c.}) \,,
\end{equation}
where $q_j \equiv (x_{j+1} - x_j)$ is the relative displacement of
neighboring particles. This hamiltonial together with the
Schr\"odinger equation $i \hbar \vec\Psi = \widehat H^{\rm e}
\vec\Psi$ generates the system of evolution equations:
\begin{equation}
 \label{k}
\left\{
 \begin{split}
\ddot x_j = & (q_j - q_{j-1}) -
  \alpha [(\psi_j^* \psi_{j+1} + {\rm c.c.}) -
          (\psi_{j-1}^* \psi_j + {\rm c.c.})]\\
\dot \psi_j = & \dfrac{i}{\widetilde h}
    \left[
    (1 - \alpha q_{j-1}) \psi_{j-1} + (1 - \alpha q_j) \psi_{j+1}
    \right],
 \end{split}
\right.
\end{equation}
where $\widetilde h$ is the dimensionless Planck's constant:
$\dfrac{\hbar}{v_0} \sqrt{\dfrac{k}{m}} \to \widetilde h \approx
0.017$.

The evolution equations in variables $q_j$ and $\psi_j$ are
\begin{equation}
 \label{kk}
\left\{
 \begin{split}
\ddot q_j = & (q_{j+1} - 2 q_j + q_{j-1}) -
            \alpha
            [ (\psi_{j+1}^* \psi_{j+2} + {\rm c.c.}) -
            2 (\psi_{j}^* \psi_{j+1} + {\rm c.c.}) +
              (\psi_{j-1}^* \psi_{j} + {\rm c.c.})
            ] \\
\dot \psi_j = & \dfrac{i}{\widetilde h}
    \left[
   (1 - \alpha q_j) \psi_{j+1} + (1 - \alpha q_{j-1}) \psi_{j-1}
    \right].
 \end{split}
\right.
\end{equation}
Eqs. \eqref{kk} are more useful for analytical consideration,
while Eqs.~\eqref{k} are more suitable for numerical calculations.

Eqs. \eqref{kk} can be reduced to the system of continuous
equations in the long-wave approximation. Let us expand the
variables $g_j$ and $\psi_j$  into Taylor series:
\begin{equation}
 \label{mm}
 \begin{split}
q_{j \pm 1}    = &  \, q_j \pm \epsilon q_j' + \epsilon^2
\dfrac12 q_j'' \pm \ldots  \\
\psi_{j \pm 1} = & \, \psi_j \pm \epsilon \psi_j' +
                    \dfrac12 \epsilon^2 \psi_j'' \pm \ldots
 \end{split}
\end{equation}
where $\epsilon$ is small parameter. After substitution of
Eqs.~\eqref{mm} into Eqs.\eqref{kk} one can get the system of
nonlinear partial differential equations (PDEs):
\begin{equation}
 \label{nn}
\left\{
 \begin{split}
 q_{tt} = & q_{xx} + 2 \alpha (\psi \psi^*)_{xx} \\
 \psi_t = & \dfrac{i}{\widetilde h}[2 (1 - \alpha q) \,
               \psi + \psi_{xx}] \,,
 \end{split}
\right.
\end{equation}
where terms of the order $\epsilon^3$ and higher are  omitted.
This system is not integrable, but it has special one-soliton
solution:
\begin{equation}
 \label{pp}
\left\{
 \begin{split}
q(x,t) = & -\dfrac{A}{\cosh^2[d(x - v_{\rm p}t)]} \\
\psi(x,t) = & \dfrac{B \, \exp[i(kx + \omega t)]}
               {\cosh[d(x - v_{\rm p}t)]} \,,
 \end{split}
\right.
\end{equation}
where $1/d$ is the polaron width and $v_{\rm p}$ is the polaron
velocity. $A, B$ are amplitude of relative displacements and
amplitude of the wave function, correspondingly, $(kx + \omega t)$
is the phase of the wave function. If $v_{\rm p} = 0$ then
\eqref{pp} coincides with the solution for the unmovable polaron
\cite{Lik12}.

Solution  \eqref{pp} is the one-parametric solution with the
following relation between parameters:
\begin{equation}
   \label{ww}
    d = \sqrt{\alpha A}; \quad
    v_{\rm p} =
    \left( 1 - \sqrt{ \dfrac{\alpha^3}{A} } \right)^{1/2}
    \approx 1-\dfrac{1}{2}\sqrt{\dfrac{\alpha^3}{A}}; \quad
    B = \sqrt{\dfrac{d}{2}} .
\end{equation}
and
\begin{equation}
   \label{www}
  k = \dfrac{\widetilde h v_{\rm p}}{2} \ll 1; \qquad
  \omega = \dfrac{2 + d^2 - k^2}{\widetilde h} \gg 1 \,.
\end{equation}
and the amplitude of relative displacements $A$ can be chosen as a
free parameter specifying the polaron. Other parameters ($d, \,
v_{\rm p}$) are defined through amplitude $A$. The norm of the
wave function $\int \limits_{-\infty}^{\infty} |\psi(x)|^2 {\rm
d}x = 1$ was preserved while deriving \eqref{ww}--\eqref{www}.

The polaron velocity $v_{\rm p}$ lies in the range from zero
(unmovable polaron) to the sound velocity (the dimensionless sound
velocity $v_{\rm snd} = 1$). The amplitude of the moving polaron
$A \geq A_{\rm min}$ and $A_{\rm min}=\alpha^3$  is the amplitude
of unmovable polaron.

\begin{figure}
 \begin{center}
  \includegraphics[width = 80 mm]{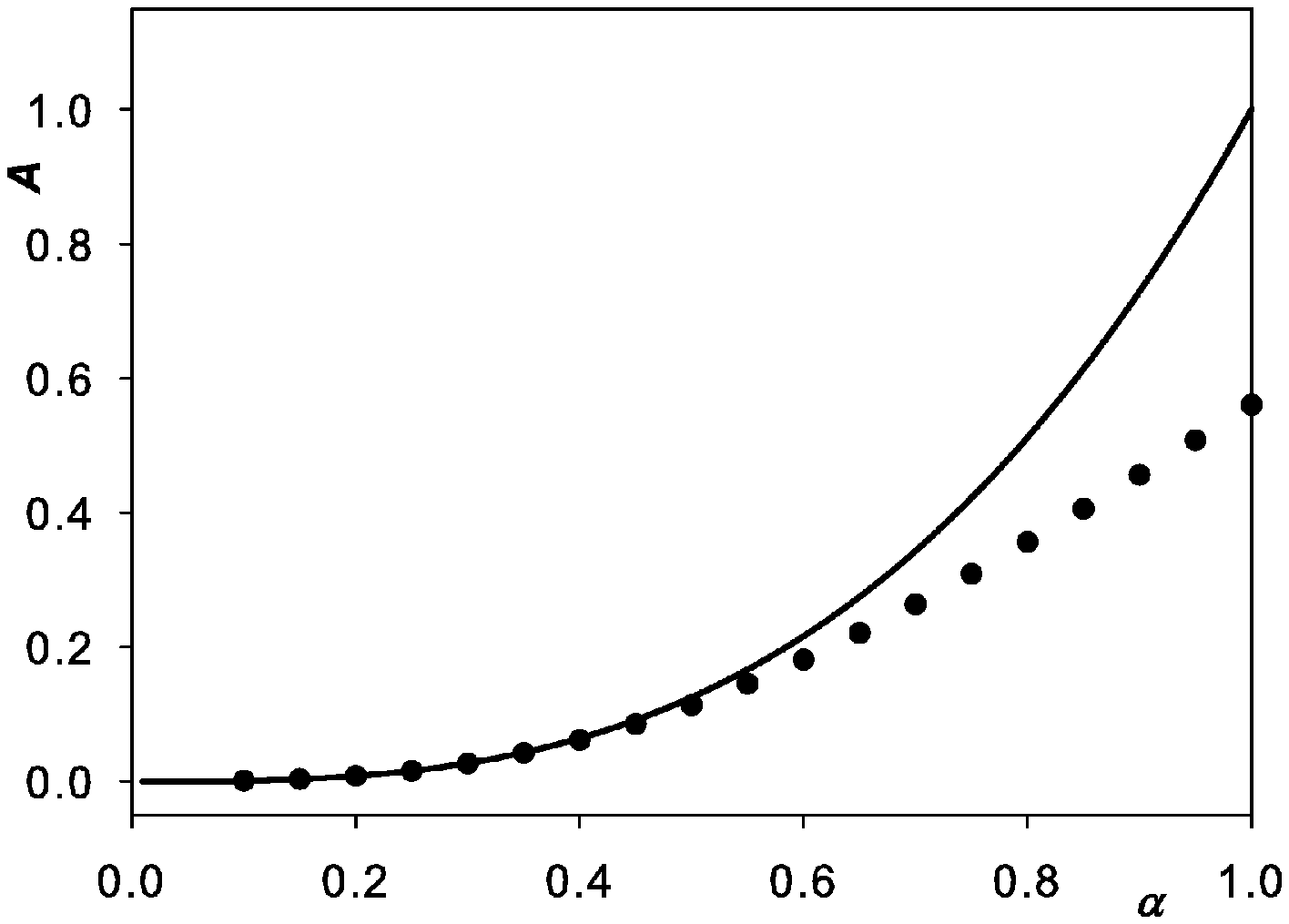}
  \includegraphics[width = 80 mm]{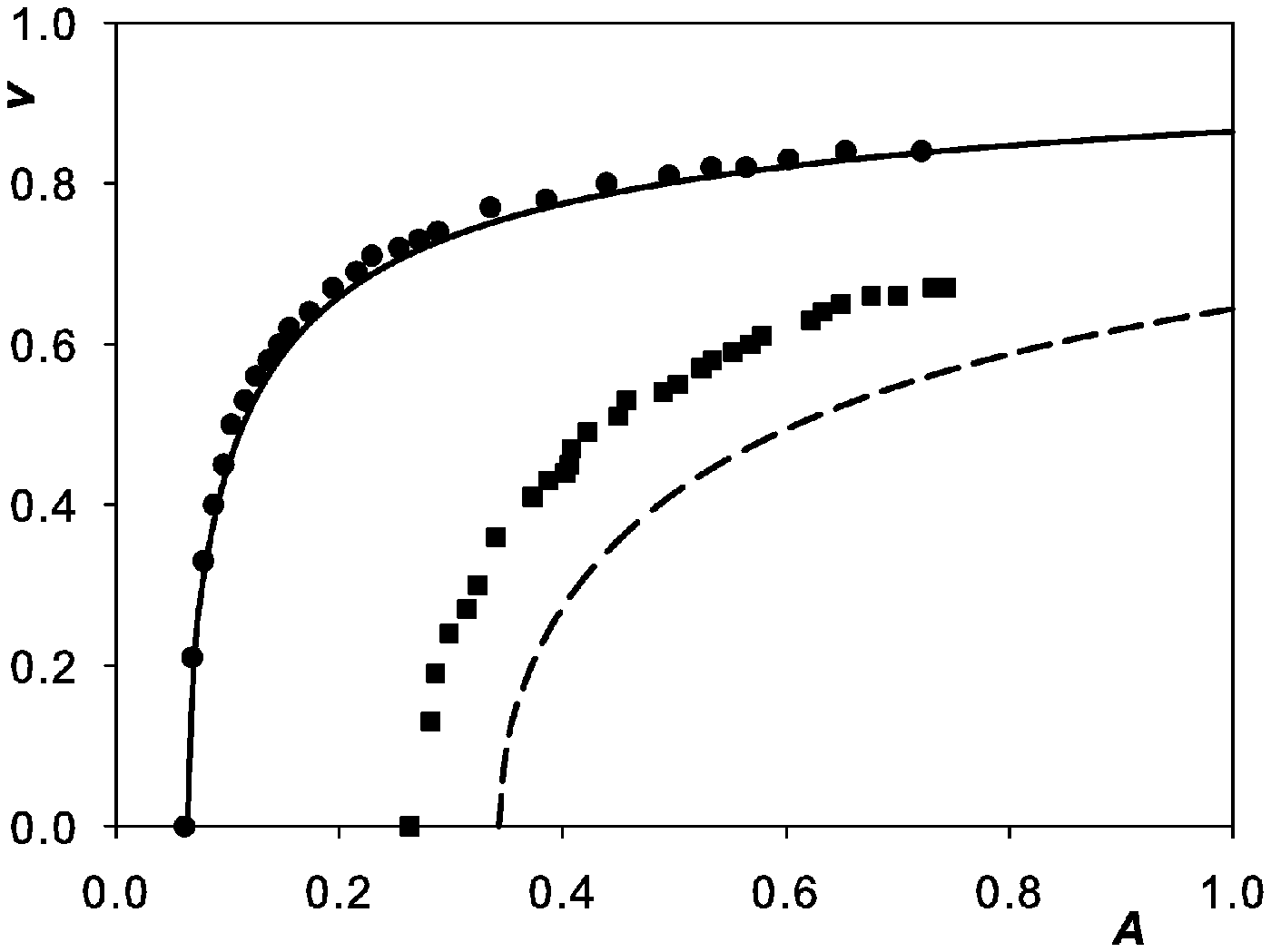}
 a) \hspace{8 cm} b)  \hspace{8 cm}
  \vspace{-0.75 cm}
  \caption{
(a) Dependence of the unmovable polaron amplitude $A$ {\it vs.}
parameter $\alpha$. Solid line is analytical prediction, dots are
results of numerical   simulation. (b) Dependence of the polaron
velocity $v$ {\it vs.} the amplitude $A$   for two values of
parameter $\alpha$.   Analytical results \eqref{ww} are shown by
solid ($\alpha = 0.4$) and dashed  ($\alpha = 0.7$) lines.
Numerical results are shown by circles ($\alpha = 0.4$) and
squares ($\alpha = 0.7$).
        }
   \label{Fig_01}
 \end{center}
\end{figure}

Now we check the analytical results in numerical simulation of
Eqs.~\eqref{k}. First of all we find the range of parameter
$\alpha$ where the continuous approximation is valid.
Fig.~\ref{Fig_01}a shows the dependence of the unmovable polaron
amplitude {\it vs.}  parameter $\alpha$ of the electron-phonon
interaction. Analytical and numerical results coincide for $\alpha
\lesssim 0{.}4$.

Fig.~\ref{Fig_01}b shows the dependence of polaron velocity {\it
vs.} its amplitude $A$ for two values of parameter $\alpha$. As
expected, coincidence is very good for $\alpha = 0.4$ and there is
some disagreement for larger values of $\alpha$. But despite some
discrepancy for $\alpha = 0{.}7$, the numerical data form a
distinct dependence. It means that there can exists the polaron
solution of \eqref{kk} which, nevertheless, is not governed by the
continuum approximation.

\begin{figure}
 \begin{center}
  \includegraphics[width=52 mm]{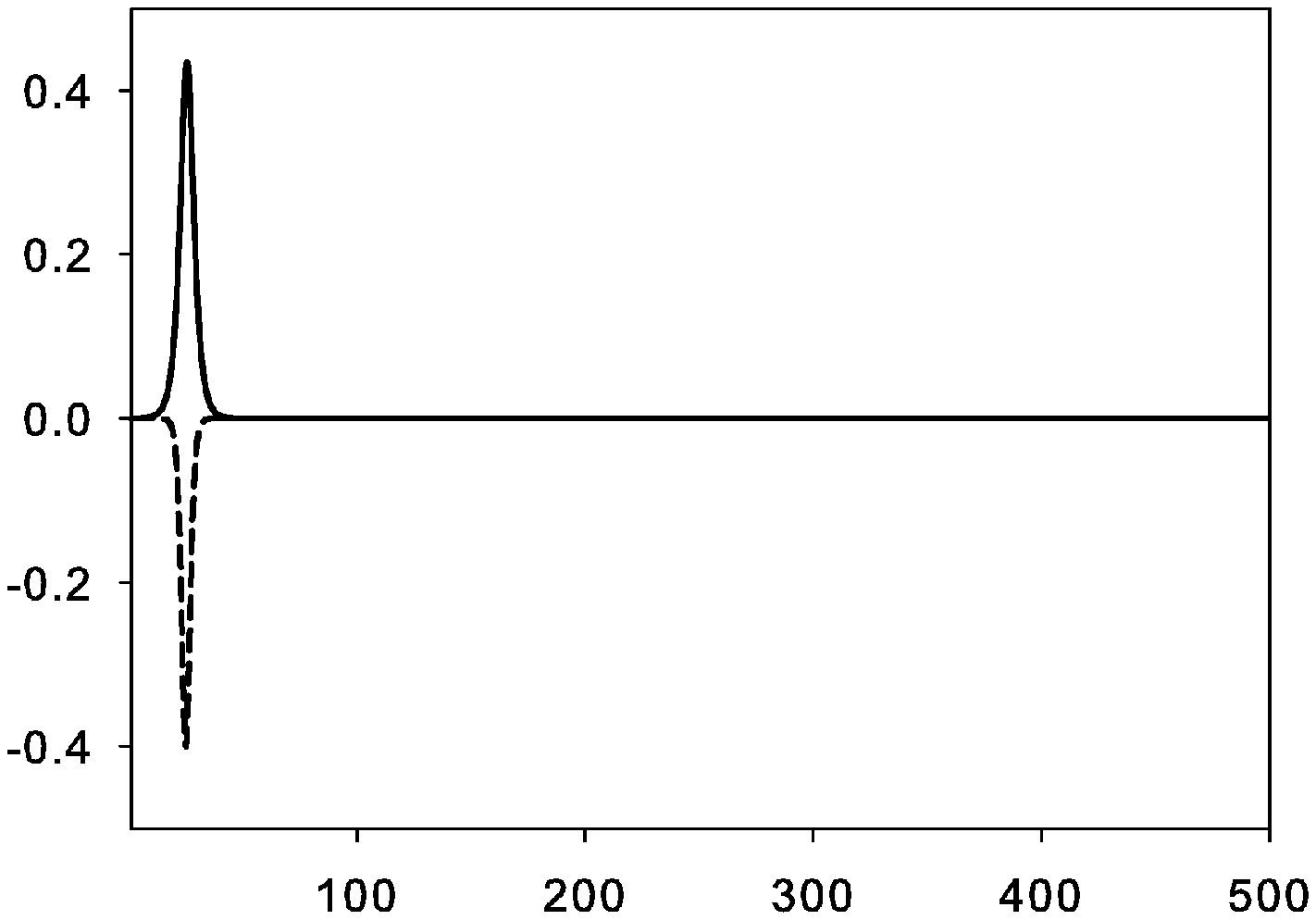}
  \includegraphics[width=52 mm]{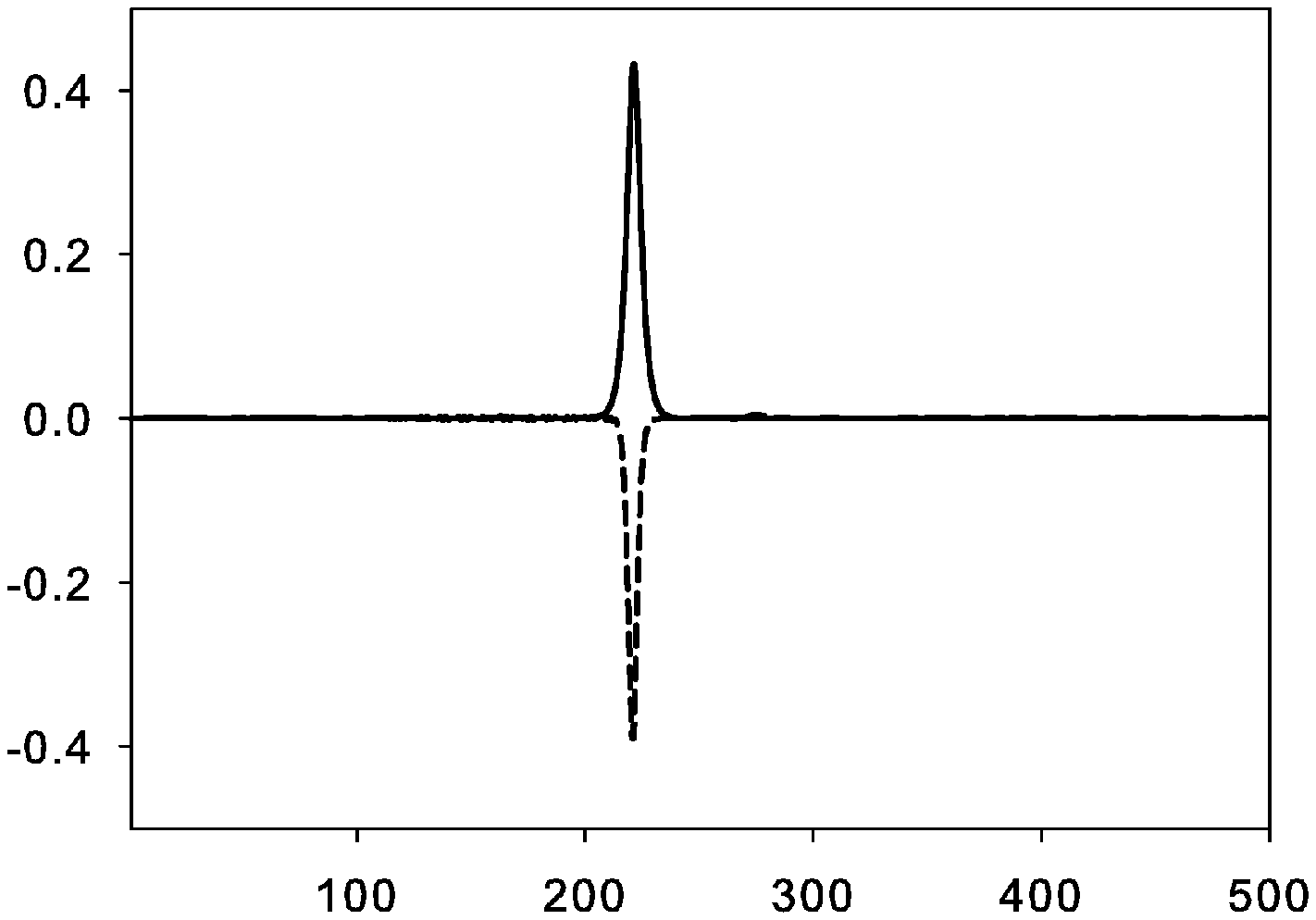}
  \includegraphics[width=52 mm]{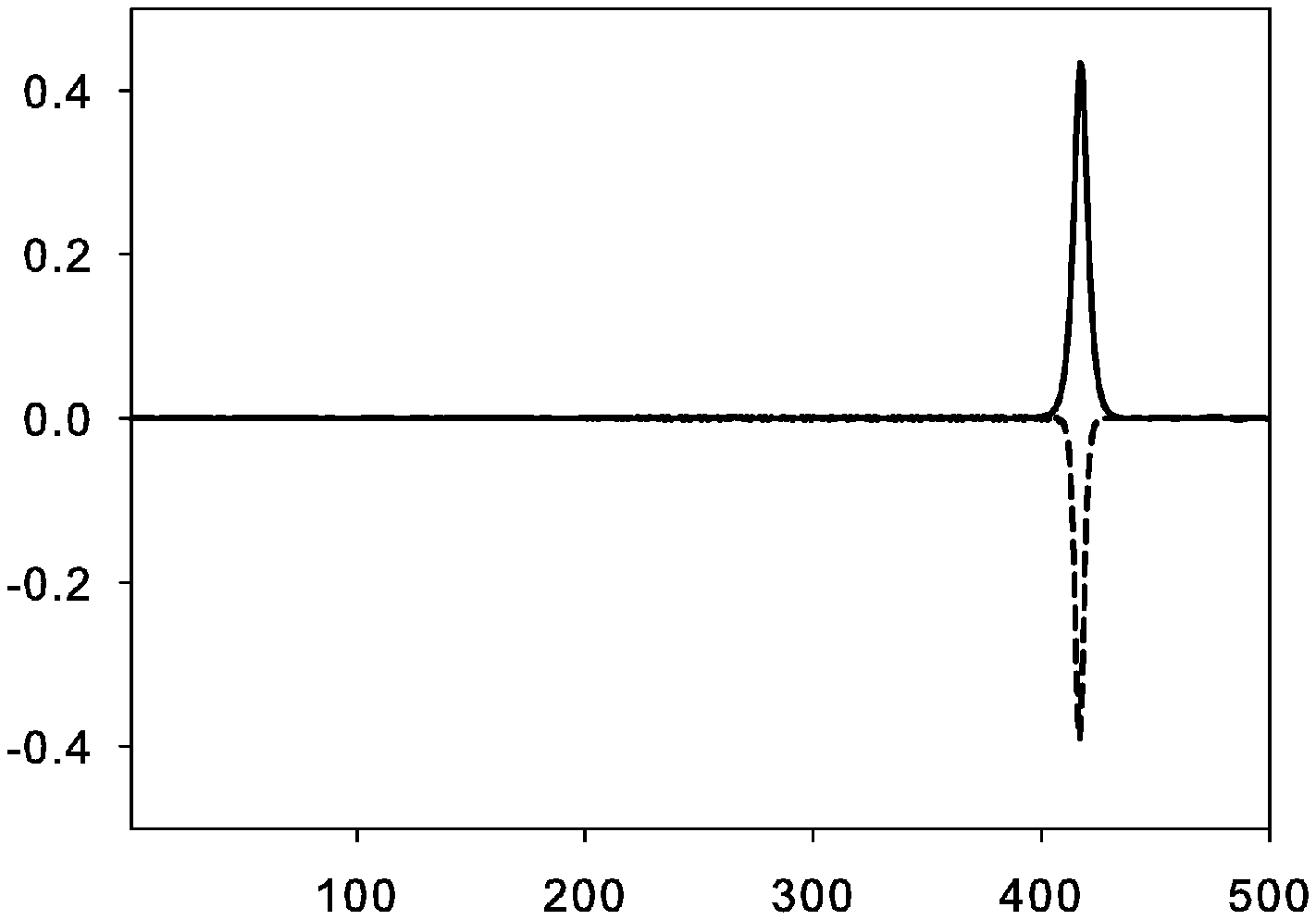}

 a) \hspace{4.5 cm} b)  \hspace{4.5 cm} c)
  \caption{
Snapshots of the polaron evolution on the harmonic lattice at
different time moments: $t=0$ (a), $t=250$ (b) and $t = 500$ (c).
Polaron is centered on the site $j_0 = 25$ at $t=0$. Positive
values along the ordinate axis are the modulus of wave function
$|\psi_j|$ (solid line), and negative values are relative
displacements (dashed line). Lattice parameters are $\alpha =
0{.}4$, $N = 500$. Initial polaron amplitude $A = 0{.}4$.
          }
   \label{Fig_02}
 \end{center}
\end{figure}

We use parameter $\alpha = 0.4$ for the analysis of the polaron
stability. Figs.~\ref{Fig_02} demonstrates high polaron stability,
where snapshots of the polaron evolution are shown at three time
moments. The initial conditions are chosen according to
\eqref{pp}-\eqref{www} for $A = 0.4$. The polaron is very stable,
it travels $\approx 400$ lattice sites without any noticeable
change. The calculated polaron velocity ($v_{\rm p} \approx 0.78$)
is in good agreement with analytical prediction ($v_{\rm anal} =
0.775$).

\begin{figure}
 \begin{center}
  \includegraphics[width=53 mm]{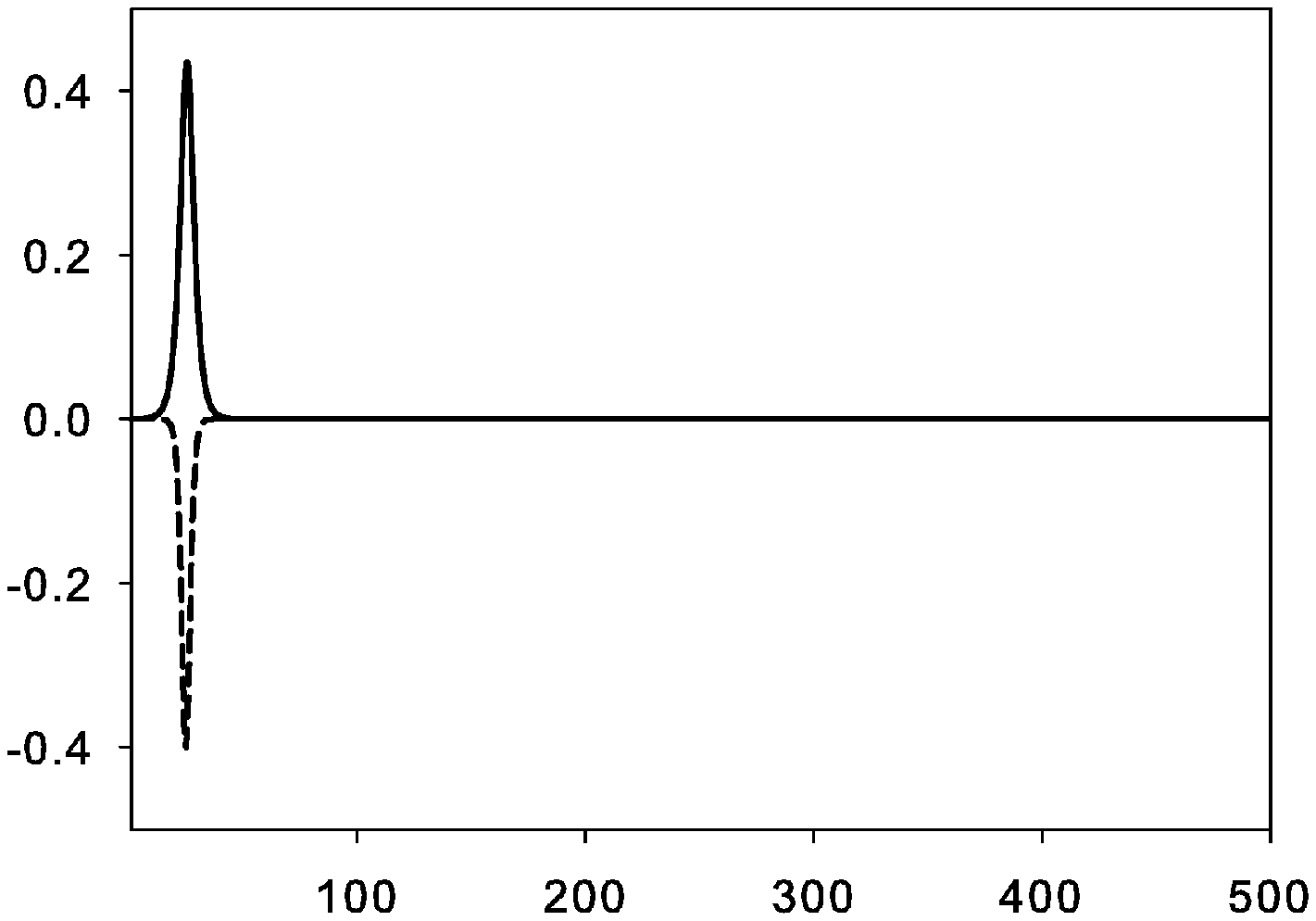}
  \includegraphics[width=53 mm]{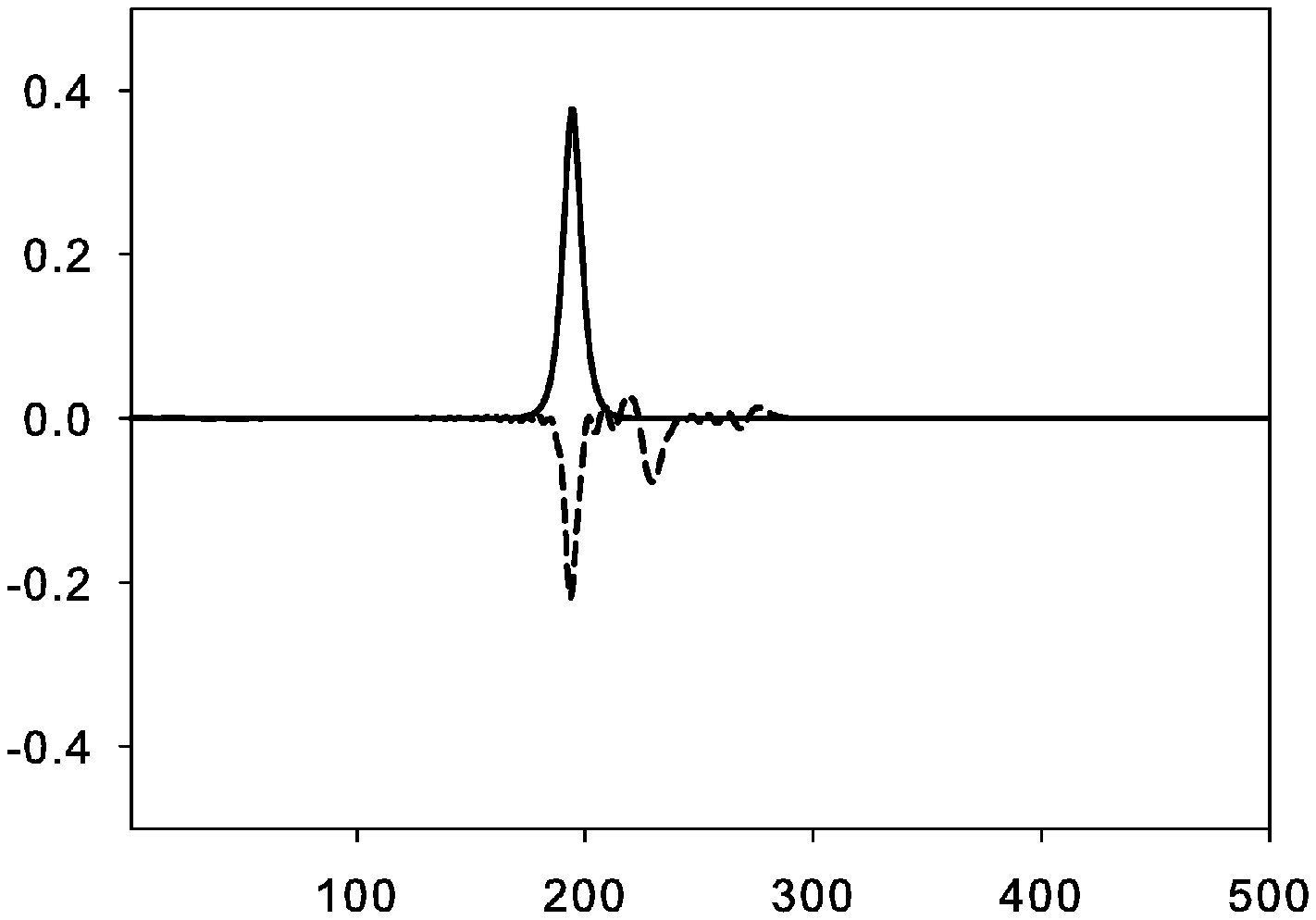}
  \includegraphics[width=53 mm]{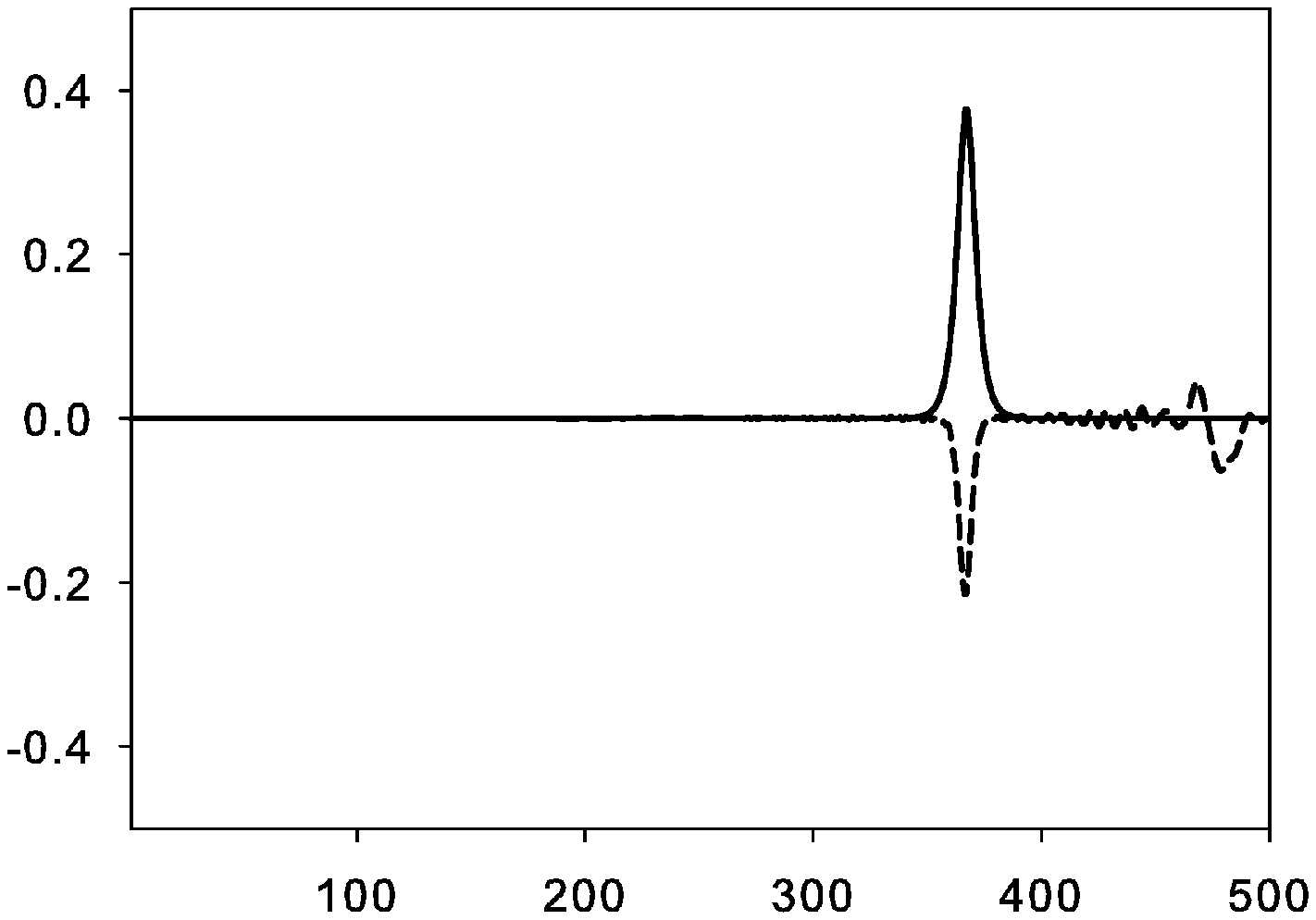}
 a) \hspace{4.5 cm} b)  \hspace{4.5 cm} c)
  \caption{
Polaron evolution with initial conditions: $\alpha = 0.4, \, A =
0.4$ and polaron velocity $v_p = 0.2$ which does not satisfy
correct relations \eqref{ww}. Snapshots are shown at $t = 0, \,\ t
= 250$ and $t = 500$.  Lines and axes are the same as in
Fig.~\ref{Fig_02}. $N = 500$.
        }
   \label{Fig_03}
 \end{center}
\end{figure}

The polaron initially generated according to the ``correct''
initial conditions is very stable. The problem is whether the
polaron can be formed and is it stable in the case of arbitrary
initial conditions? To answer this question the initial excitation
is chosen according to \eqref{pp}-\eqref{www} for $A = 0{.}4$, but
the velocity is $v_{\rm p} = 0{.}2$ instead of correct value
$v_{\rm anal} = 0{.}775$. The evolution of this initial state is
shown in Fig.~\ref{Fig_03}. Stable polaron with parameters $A
\approx 0{.}21, \,\, v_{\rm p} \approx 0{.}68$ forms soon
(polarons in Figs. \ref{Fig_03}b and \ref{Fig_03}c are practically
identical). The polaron amplitude decreased relative to initial
value. But relations \eqref{ww} and \eqref{www} are fulfilled for
parameters of this polaron with high accuracy, i.e. the analytical
polaron velocity $v_{\rm anal} \approx 0.67$ for the polaron
amplitude $A=0.21$. The initial excitation transforms into polaron
accompanied by the noise irradiation. Noteworthy, that 100\% of
the electron wave function is localized at polaron; no fraction of
the wave function is irradiated with the noise.

Wide range of initial conditions was tested and in all cases
stable polaron is formed with the relations between parameters
obeying \eqref{ww}--\eqref{www}.

In the next section we transform Eqs.~\eqref{nn} into the
integrable system of PDEs.


\section{Reduction to the integrable system of partial \\
differential equations.}

Let us rewrite system \eqref{nn} for convenience:
\begin{equation}
 \label{nn1}
\left\{
 \begin{split}
 q_{tt} = & q_{xx} + 2 \alpha (\psi \psi^*)_{xx} \\
 \psi_t = & \dfrac{i}{\widetilde h}[2 (1 - \alpha q) \,
               \psi + \psi_{xx}] \,.
 \end{split}
\right.
\end{equation}

We follow  the {\it reductive perturbation method} (RPM)
\cite{Sas81, Leb08} where a special variable transformation
\begin{equation}
 \label{RPM}
\left\{
 \begin{split}
  q    = \, & \varepsilon u_1 + \varepsilon^2 u_2 + \ldots \\
  y    = \, & \varepsilon^{1/2} (x-c \,t) \\
  \tau = \, & \varepsilon^{3/2} t ,
 \end{split}
\right.
\end{equation}
is made and $\varepsilon$ is small parameter. The main objective
of the RPM usage is the reduction of the degree of derivative by
time $q_{tt}$ in \eqref{nn1} to the first order. The new
(laboratory) coordinate system moves with velocity $c$ relative to
the old coordinate system (second line in \eqref{RPM} ).

Substituting \eqref{RPM} into \eqref{nn1} and equating terms of
the order $\varepsilon^2$ one gets that  $c = 1$. It means that
the laboratory coordinate system moves with the sound velocity $c=
v_{\rm snd} = 1$ relative to the old coordinate system.

Equating terms of the order $\epsilon^3$ and taking into account
that the solution tends to zero at $\pm \infty$, we get
\begin{equation}
 \label{Eq1}
\left\{
 \begin{split}
{} &  2 u_{\tau} = 2 \alpha (\psi \psi^*)_y   \\
{} &  i \widetilde{h} (\psi_{\tau} - \psi_y) =
      -2 (1 - \alpha u) \psi - \psi_{yy}.
 \end{split}
\right.
\end{equation}
Additional linear variables transformation ($\psi(y,\tau) =
\widetilde{a} \, \Psi(z,T) \, \exp[i(\widetilde kz - \widetilde
\omega T)], \,\, u(y,\tau) = \widetilde{b} \, U(z,T), \,\, \tau =
\widetilde{c} \, T, \,\, y = \widetilde d z$) is necessary to
reduce \eqref{Eq1} to the ``standard'' form:
\begin{equation}
 \label{Eq6}
\left\{
 \begin{split}
  U_T      = &  \, 2 (\Psi \Psi^*)_z   \\
  i \Psi_T = & \, U \Psi + \Psi_{zz},
 \end{split}
\right.
\end{equation}
where coefficients labelled with `tilde' are some linear
combinations of ``old'' coefficients.

System  \eqref{Eq6} is exactly solvable and its one-, two-soliton
etc. solutions are known \cite{Dub88}. As a particular case, the
one-soliton solution has the same functional form as solution
\eqref{pp}. But the relation between parameters $A, \, B, \, d, \,
v_{\rm p}$ is somewhat different. Numerical simulation confirms
high stability of polarons described by the solution of
\eqref{Eq6}.

The numerical experiments of polaron interaction shows that the
polarons preserve their shapes after collision. This fact
additionally proves the fact that integrable system \eqref{Eq6} is
a good approximation for discreet system \eqref{kk}.

In the next section we consider the moving polaron on the
inharmonic lattice.


\section{Moving polaron on the inharmonic lattice}

In this section we consider the inharmonic lattice with the
dimensionless FPU potential:
\begin{equation}
   \label{Morse3}
    U(q) = \dfrac12 q^2 - \dfrac{\beta}{3} q^3,
\end{equation}
where  $\beta$ is the parameter of non-linearity and $q$ is the
relative displacement of particles. Many realistic potentials
(Morse, Lennard-Jones, etc) are reduced to \eqref{Morse3} when the
deviations from equilibrium are not large (the expansion of
potentials into the Tayler series up to the third order).
Moreover, the $\alpha$-FPU potential \eqref{Morse3} allows to make
necessary analytical calculus.

We employ the same procedure as in the case of harmonic potential.
We expand the relative displacements and the wave function into
Taylor series according to \eqref{mm}. The result is the system of
PDEs:
\begin{equation}
   \label{a1}
 \left\{
 \begin{split}
  q_{tt}  = &
     \left(q_{xx} + \dfrac{1}{12} q_{xxxx} \right) -
     \beta \left( q^2 \right)_{xx} +
     2 \alpha \left( \psi \psi^* \right)_{xx}; \\
  \psi_t = &
   \dfrac{i}{\widetilde h}[2 (1 - \alpha q) \,
               \psi - \psi_{xx}].
 \end{split}
\right.
\end{equation}
Unfortunately, the relation between coefficients in \eqref{a1} is
such, that this system can not be reduced to the exactly solvable
PDEs \cite{Dub88} which has the canonical form:

\begin{equation}
   \label{a19}
 \left\{
 \begin{split}
  3 u_{tt} =  & (u_{xxx}-6 u \, u_x)_x + 8 |\phi|^2_{xx}; \\
    i \, \phi _t = & \phi_{xx}-u \phi.  \\
 \end{split}
\right.
\end{equation}

However, if $\beta = 2 \, \alpha$ (coefficients of the second
spatial derivatives in the first equation in \eqref{a1} are
equal), then system \eqref{a1} has the special one-soliton
solution. This solution has the same form as for the harmonic
lattice:
\begin{equation}
 \label{pp1}
\left\{
 \begin{split}
q(x,t) = & -\dfrac{A}{\cosh^2[d(x - vt)]} \\
\psi(x,t) = & \dfrac{B \, \exp[i(kx + \omega t)]}
               {\cosh[d(x - vt)]}
 \end{split}
\right.
\end{equation}
with the relations between parameters:
\begin{equation}
 \label{rel1}
  \begin{array}{l}
  d = \sqrt{\alpha A} = \sqrt{2 \beta A}, \\
  v_{\rm p} = \left( 1 -
  \sqrt{\dfrac{\alpha^3}{A}}  + \dfrac{\beta A}{3} \right)^{1/2}
  \approx   1 -  \dfrac{1}{2}\sqrt{\dfrac{\alpha^3}{A}}  +
  \dfrac{\beta A}{6}, \\
  B = \sqrt{\dfrac{d}{2}}
  \end{array}
\end{equation}
and
\begin{equation}
   \label{rel2}
  k = \dfrac{\widetilde h v_{\rm p}}{2} \ll 1; \qquad
  \omega = \dfrac{2 + d^2 - k^2}{\widetilde h} \gg 1.
\end{equation}
There is an addition term in the expression for the polaron
velocity \eqref{rel1} as compared to the polaron velocity on the
harmonic lattice \eqref{ww}. This positive term is due to the
nonlinearity, and the actual velocity is determined by the balance
of electron-phonone interaction and nonlinearity. In the absence
of the electron-phonon interaction ($\alpha = 0$) the velocity is
$v \approx 1 + \beta A/6$, and is the soliton velocity on the
$\alpha$-FPU lattice \cite{Rem03}.

\begin{figure}
 \begin{center}
  \includegraphics[width=80 mm]{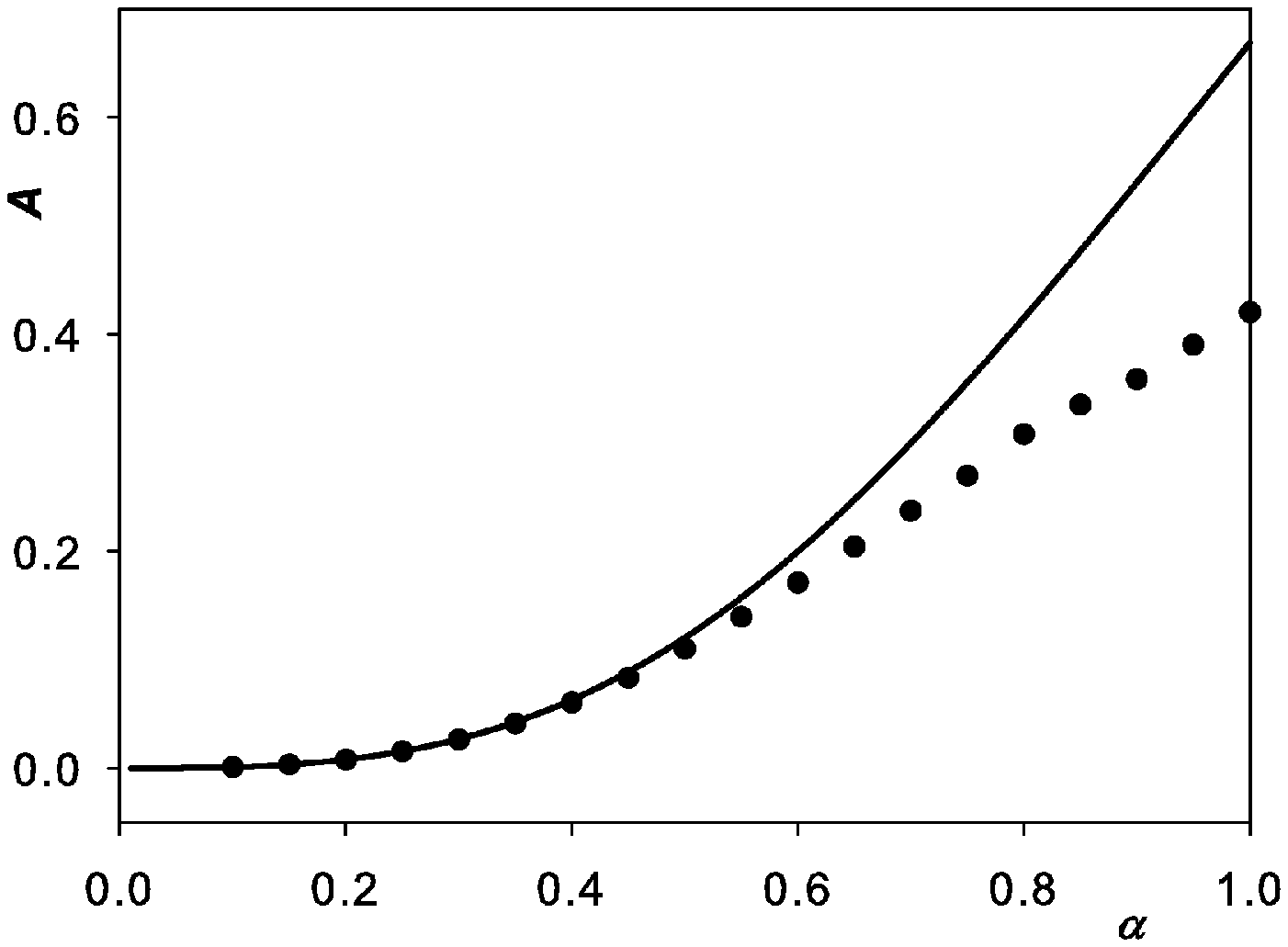}
  \includegraphics[width=80 mm]{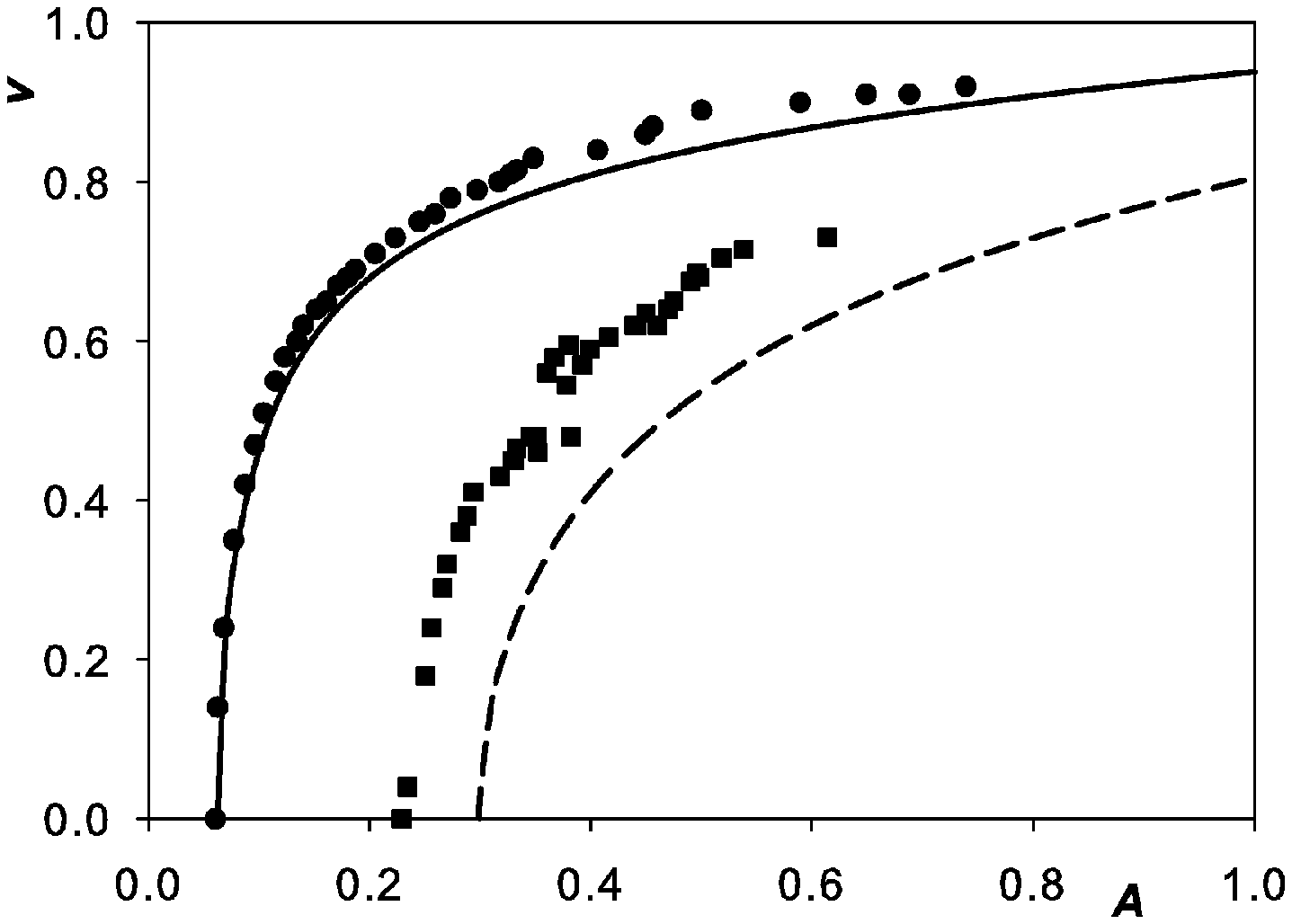}
 a) \hspace{8 cm} b)  \hspace{8 cm}
  \vspace{-0.75 cm}
 \caption{
(a) The dependence of the polaron amplitude {\it vs.} parameter
$\alpha$ for the unmovable polaron. Solid line is analytical
solution \eqref{pp1} with parameters \eqref{rel1}--\eqref{rel2},
dots are numerical results. (b) Dependence of the polaron velocity
{\it vs.} amplitude $A$ for two values of parameter $\alpha$:
$\alpha = 0{.}4$ (solid line is analytical dependence, circles are
numerical results), $\alpha = 0{.}7$ (dashed line is analytical
dependence, squares are numerical results).
        }
   \label{Fig_04}
 \end{center}
\end{figure}

Now we verify the polaron stability in numerical simulations on
the inharmonic lattice in a way analogous to the harmonic lattice.
Fig.~\ref{Fig_04}a shows the dependence of the unmovable polaron
amplitude {\it vs.} the parameter $\alpha$ of the electron-phonon
interaction. One can see, that there is also a good agreement
between analytical and numerical results for $\alpha \lesssim
0{.}4$.

Figure \ref{Fig_04}b shows the dependence of the polaron velocity
{\it vs.} amplitude $A$ of relative displacements for two value of
parameter of electron-phonon interaction $\alpha$. A good
agreement between analytical and numerical results is also
observed for $\alpha = 0{.}4$ but there is some gap for $\alpha =
0{.}7$. Similar to the harmonic lattice, the numerical data form a
well-defined dependence for $\alpha = 0{.}7$. It means that there
can exist polarons which are not governed by the continuum
approximation.

\begin{figure}
 \begin{center}
  \includegraphics[width=52 mm]{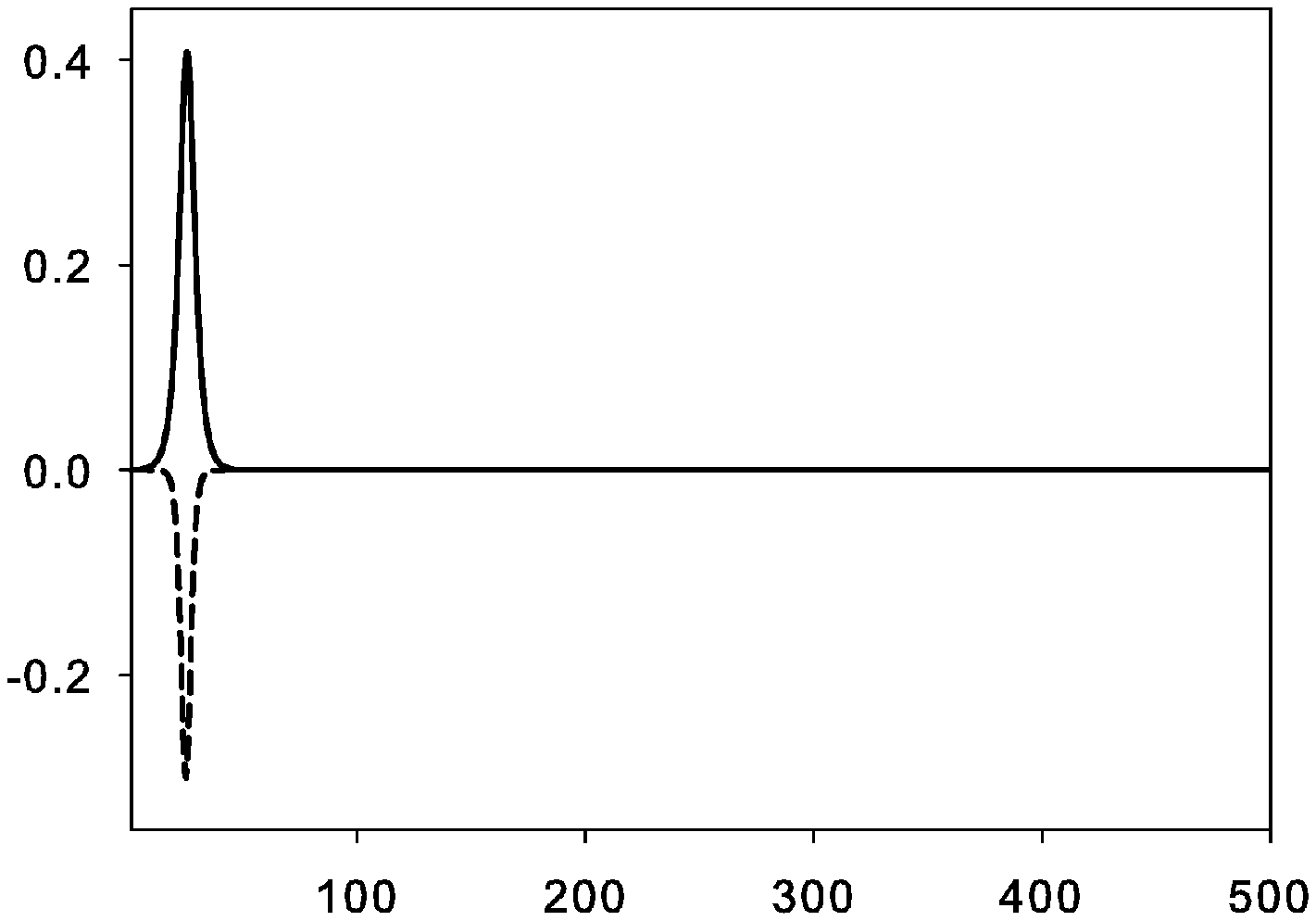}
  \includegraphics[width=52 mm]{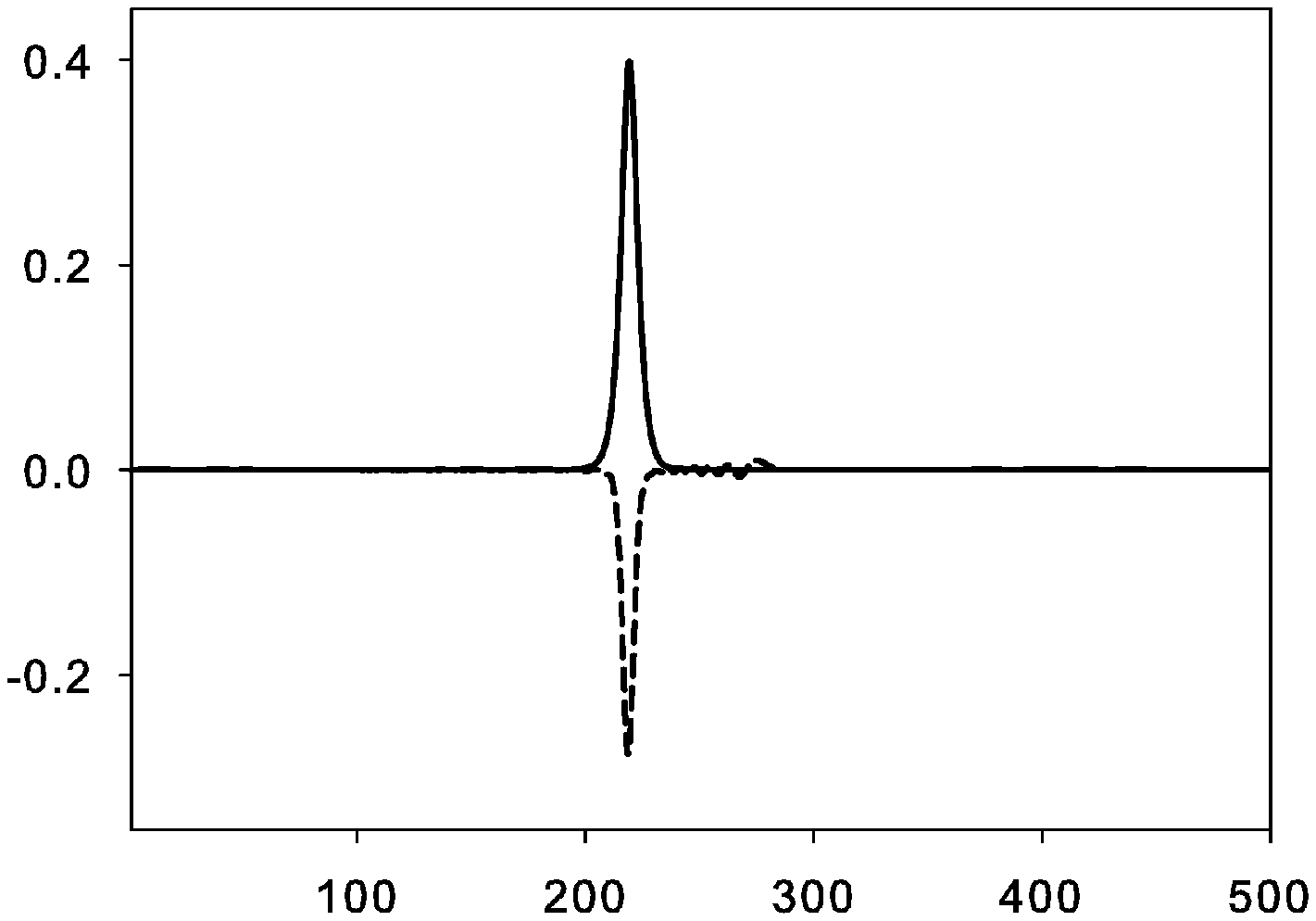}
  \includegraphics[width=52 mm]{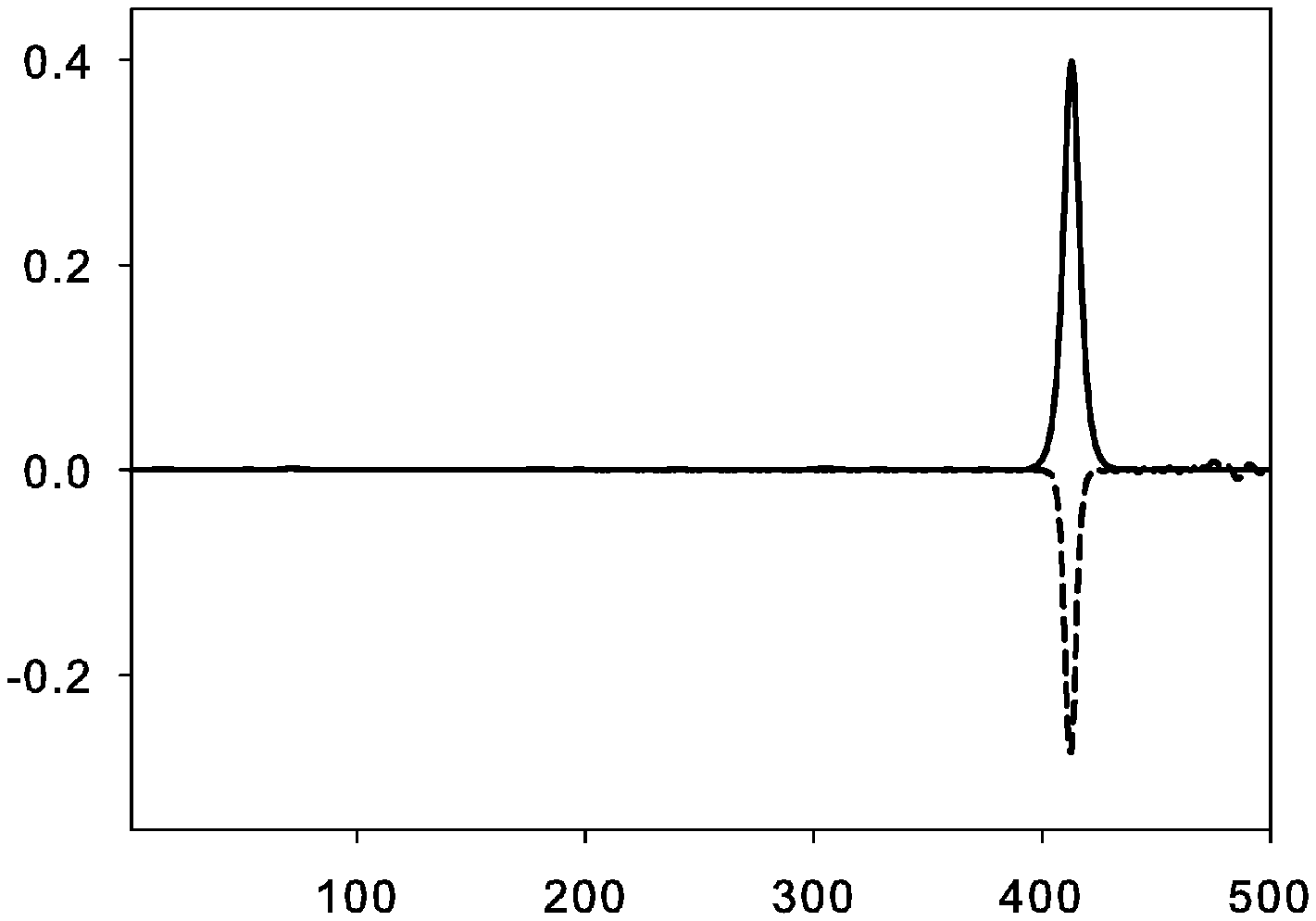}

 a) \hspace{4.5 cm} b)  \hspace{4.5 cm} c)
  \caption{
Snapshots of the polaron evolution on inharmonic lattice ($\alpha
=  0{.}4$, $\beta = 0.2$) at different time moments: $t=0$ (a),
$t=250$ (b) and $t = 500$ (c). The polaron is centered on the site
$j_0 = 25$ at $t=0$, the initial polaron amplitude $A = 0{.}3$.
Positive values along the ordinate axes are modulus of the wave
function $|\psi_j|$ (solid line), negative values are relative
displacements $q_j$ (dashed line).
  }
   \label{Fig_05}
 \end{center}
\end{figure}

Numerical simulations show high polaron stability.
Figs.~\ref{Fig_05} illustrates snapshots of the polaron evolution
at three time moments, when the initial conditions are generated
according to \eqref{pp1}--\eqref{rel2}. The initial polaron
amplitude is $A=0{.}3$. The polaron preserves initial parameters
in the course of the evolution. The calculated polaron velocity is
$v_{\rm p} \approx 0{.}78$, whereas the analytical velocity
$v_{\rm anal} = 0{.}76$ for $A=0{.}3$.

\begin{figure}
 \begin{center}
  \includegraphics[width=52 mm]{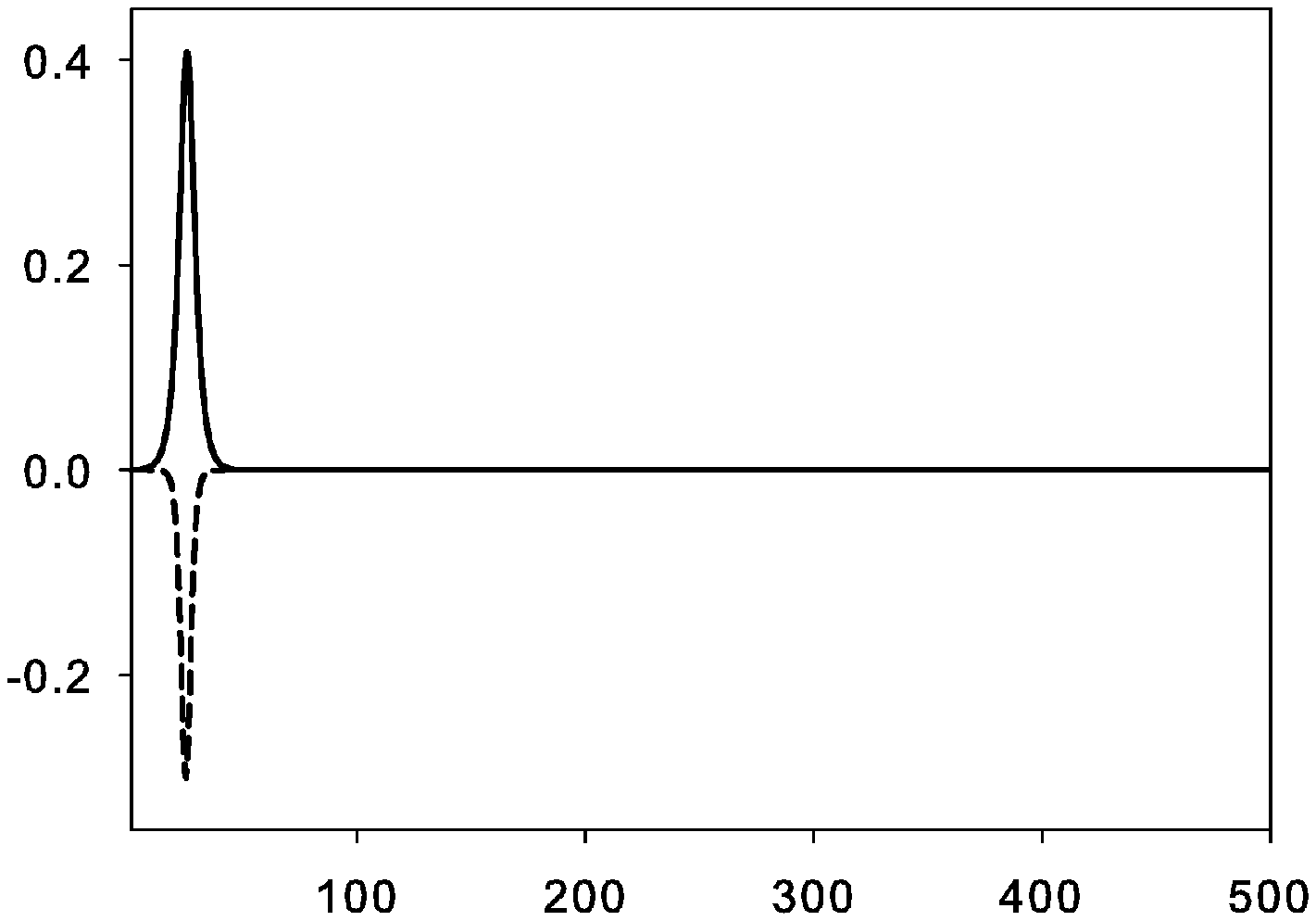}
  \includegraphics[width=52 mm]{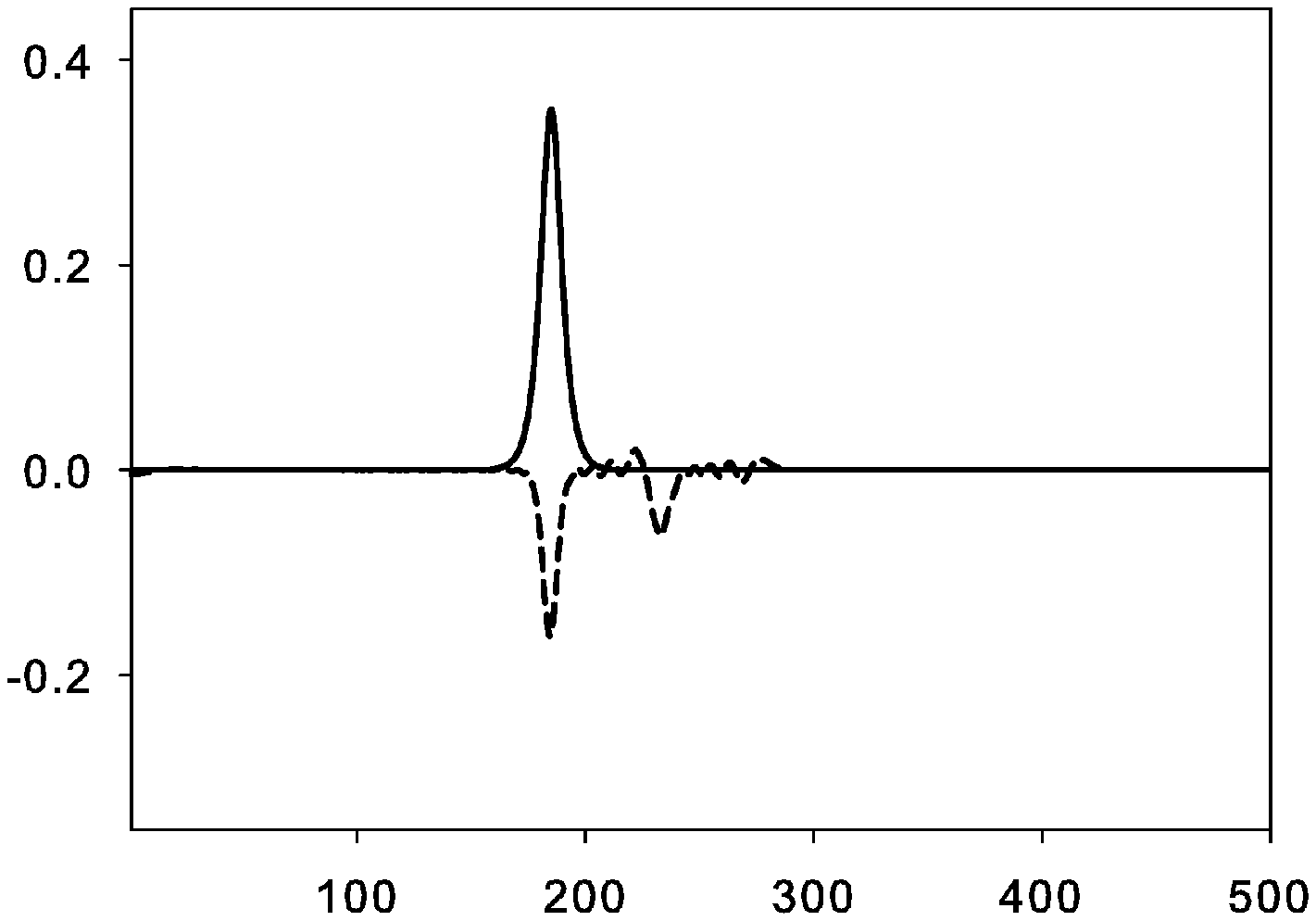}
  \includegraphics[width=52 mm]{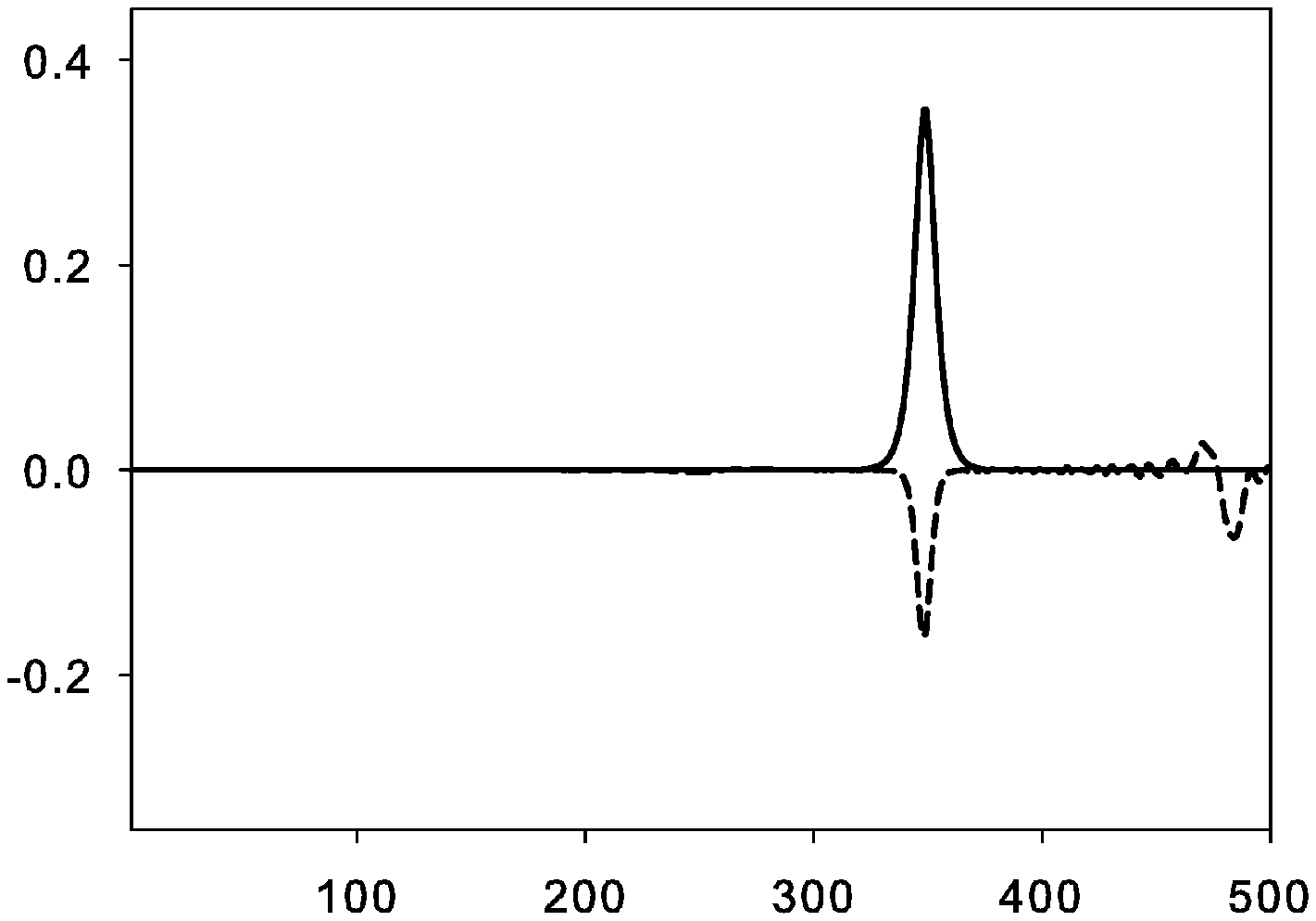}

 a) \hspace{4.5 cm} b)  \hspace{4.5 cm} c)
  \caption{
Polaron evolution with improper initial conditions. Snapshots are
shown at $t=0$ (a), $t=250$ (b) and $t = 500$ (c). Notation and
axes are the same as in Fig.~\ref{Fig_05}
        }
   \label{Fig_06}
 \end{center}
\end{figure}

The polaron evolution with improper initial conditions is shown in
Figs.~\ref{Fig_06}. The initial polaron is formed according to
Eqs.\eqref{pp1}-\eqref{rel2}, but the polaron velocity is chosen
to be $v_{\rm p} = 0{.}2$ instead of analytically predicted
$v_{\rm anal} = 0{.}76$. The simulation show that the polaron
self-organizes (compare polarons in Figs. \ref{Fig_06}b and
\ref{Fig_06}c). Parameters of self-organized polaron satisfy
\eqref{rel1}--\eqref{rel2} with high accuracy. The ``numerical''
polaron has velocity and amplitude $v_{\rm p} \approx 0{.}65$ and
$ A \approx 0{.}16$ whereas $v_{\rm anal} = 0{.}63$ if $A=0{.}16$
for the ``analytical'' polaron.

The numerical simulations show that polaron  is stable  on the
inharmonic lattice. However, the results demonstrated above are
valid only if $\alpha = 2 \, \beta$ and $\alpha \lesssim 0{.}4$.
In this case the lattice inharmonicity is small ($\beta=0{.}2$)
and does not influences essentially on the polaron evolution. The
case of large $\alpha$ and $\beta$ are of greater interest. This
case is considered numerically in the next Section.


\section{Numerical simulation on the nonlinear lattice \\
with arbitrary parameters}

Equations
\begin{equation}
 \label{k_nl}
\left\{
 \begin{split}
\ddot x_j = & (q_j - q_{j-1})
 + \beta(q_{j-1}^2 - q_j^2)
 - \alpha [(\psi_j^* \psi_{j+1} + {\rm c.c.}) -
          (\psi_{j-1}^* \psi_j + {\rm c.c.})], \\
\dot \psi_j = & \dfrac{i}{\widetilde h}
    \left[
    (1 - \alpha q_{j-1}) \psi_{j-1} + (1 - \alpha q_j) \psi_{j+1}
    \right]; \qquad q_j \equiv x_{j+1} - x_j
 \end{split}
\right.
\end{equation}
are integrated numerically with different initial conditions. We
use initial conditions in the form \eqref{pp1}, but $A$, $d$, $v$
are arbitrary parameters and $k=0$. The question is whether
polarons are really exist in the case when the continuous
approximation is not valid.

\begin{figure}
 \begin{center}
  \includegraphics[width=80 mm]{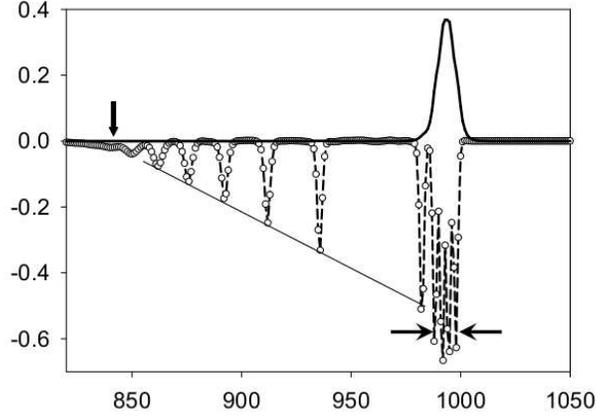}
  \caption{
The result of the evolution of the initial excitation at $t=800$.
The formed polaron is indicated by two arrows for clarity. There
are six solitons behind the polaron. The front of the sound
propagation is shown by the vertical arrow. The initial excitation
is centered at $j_0 = 50$. Modulus of the wave function -- solid
line; relative lattice displacements -- dashed line. $N = 1200$.
           }
   \label{Fig_07}
 \end{center}
\end{figure}

Fig.~\ref{Fig_07} illustrates how the strange four-peaked
polaron-like excitation self-organizes on the inharmonic lattice
with parameters  $\alpha = 0.4$, $\beta = 1.0$. The polaronic
nature of this excitation is supported by the 100\% bearing of the
wave function by the localized lattice compression. Initial
excitation at $t = 0$ is chosen with parameters $d = 0.1$, $A =
0.6$, $v_{\rm p} = 0.7$, $k=0$. Lattice excitations behind the
polaron do not have the electronic components and are solitons.
The linear relation between soliton velocities and amplitudes
\cite{Rem03} is clearly visible in Fig.~\ref{Fig_07} and
additionally supports the solitonic nature of these excitations.
This numerical simulation is an example of how an arbitrary
initial conditions transform to the polaron with extra
perturbations emitted in the forms of polaron. Noteworthy that
there are no fluctuating lattice vibrations.

\begin{figure}
 \begin{center}
  \includegraphics[width=53 mm]{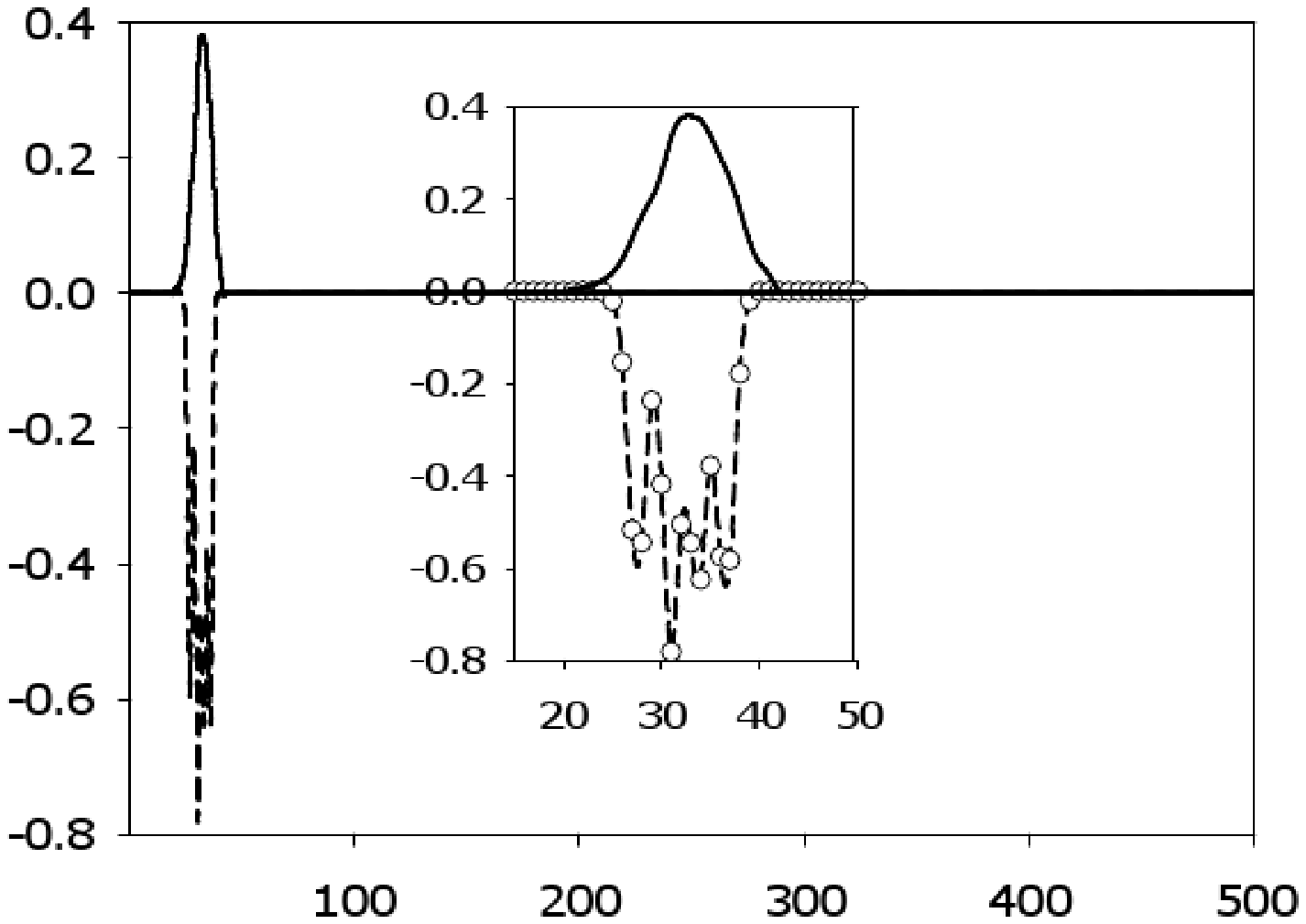}
  \includegraphics[width=53 mm]{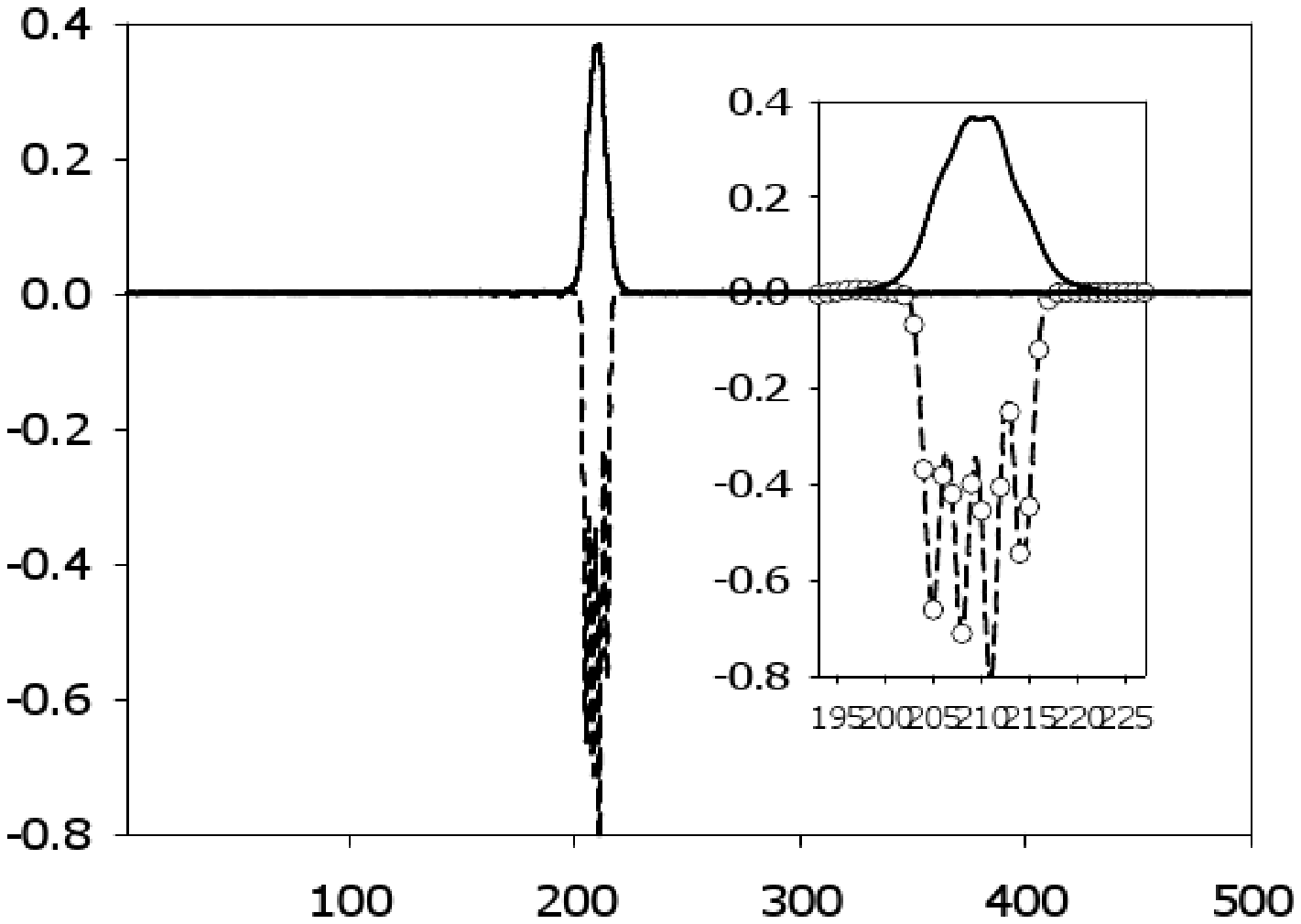}
  \includegraphics[width=53 mm]{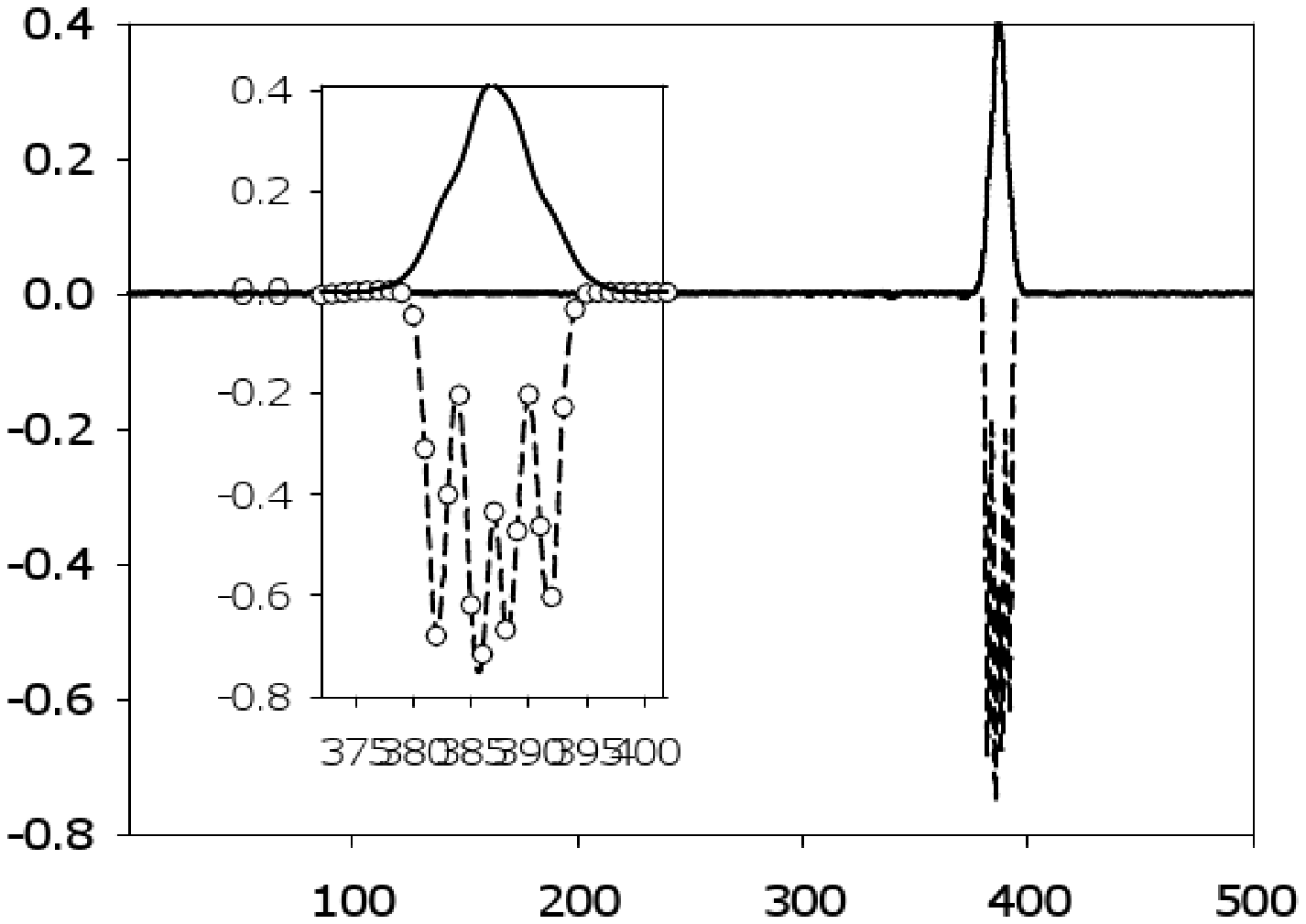}

 a) \hspace{4.5 cm} b)  \hspace{4.5 cm} c)

  \caption{
The evolution of the four-peaked polaron formed from the initial
conditions of previous numerical experiment. Inserts show polarons
with higher resolution. $N=500$.
          }
   \label{Fig_08}
 \end{center}
\end{figure}

To verify the stability of the obtained four-peaked polaron, we
use it parameters ($x_j, \, v_j, \, \psi_j; \, j = 980 \div
1020$), located between two horizontal arrows in
Fig.~\ref{Fig_07}, as the initial condition for new numerical
simulation. The result of evolution is shown in Fig.~\ref{Fig_08}.
The polaron is very stable. After travelling through $\approx \!
400$ lattice sites its parameters are not changed and it moves
with constant supersonic velocity $v_{\rm p} \approx 1{.}14$.

The most natural and expected shape of the polaron envelope is a
smooth bell-shaped form and an existence of four-peaked polaron is
quite unexpected. But there exist polarons with other number of
peaks.

\begin{figure}
 \begin{center}
  \includegraphics[width=80 mm]{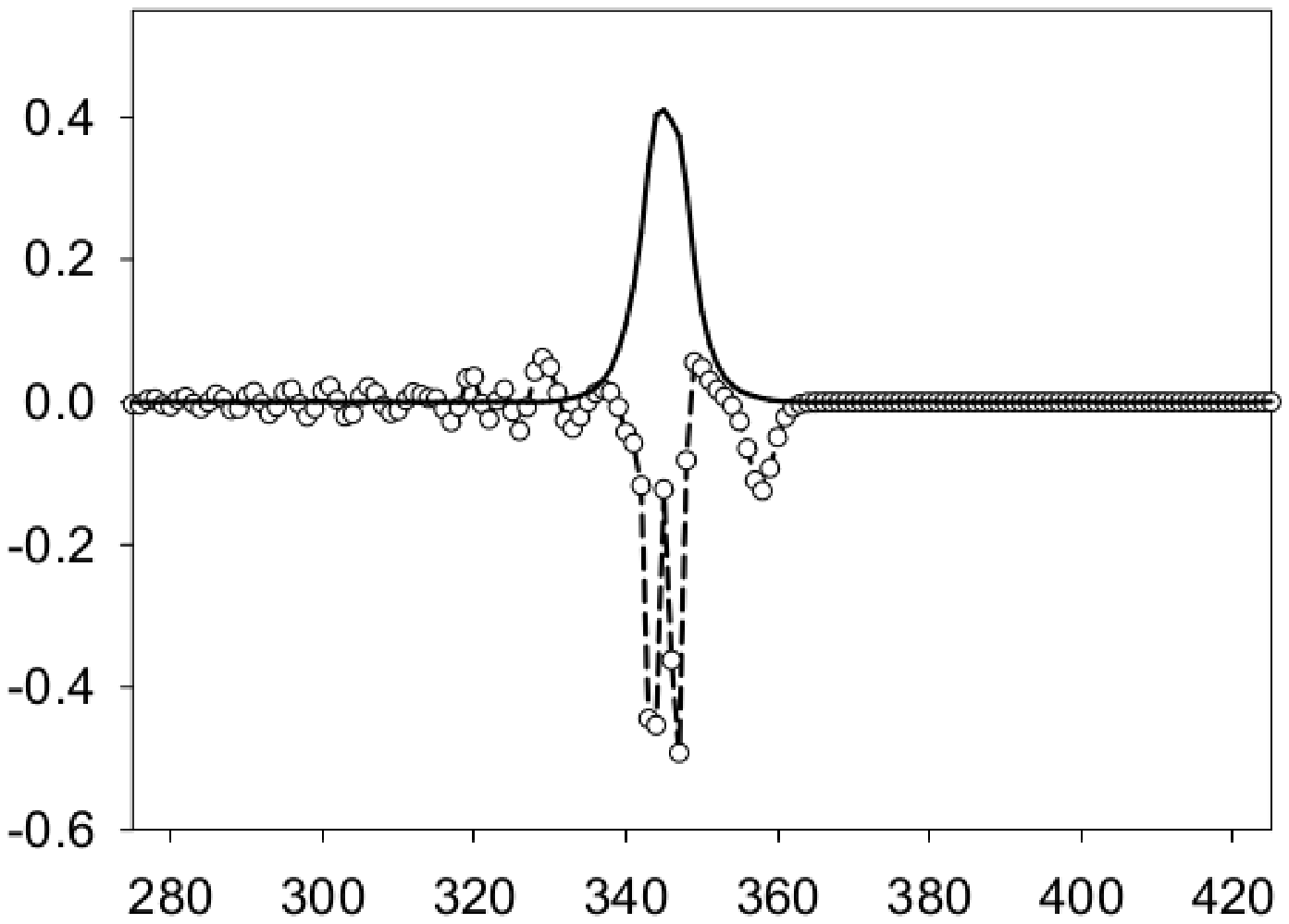}
  \includegraphics[width=80 mm]{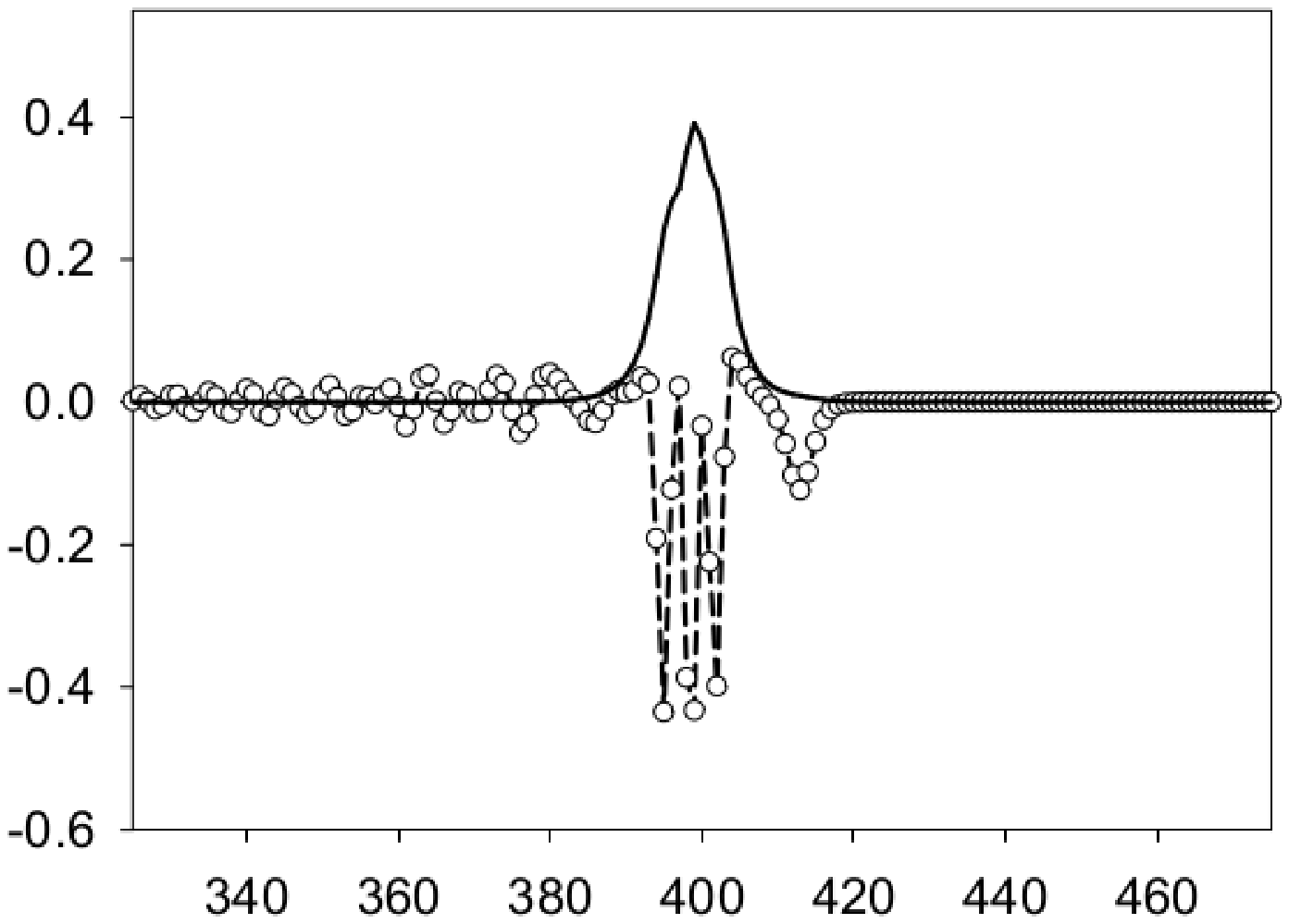}
 a) \hspace{8 cm} b)
  \includegraphics[width=80 mm]{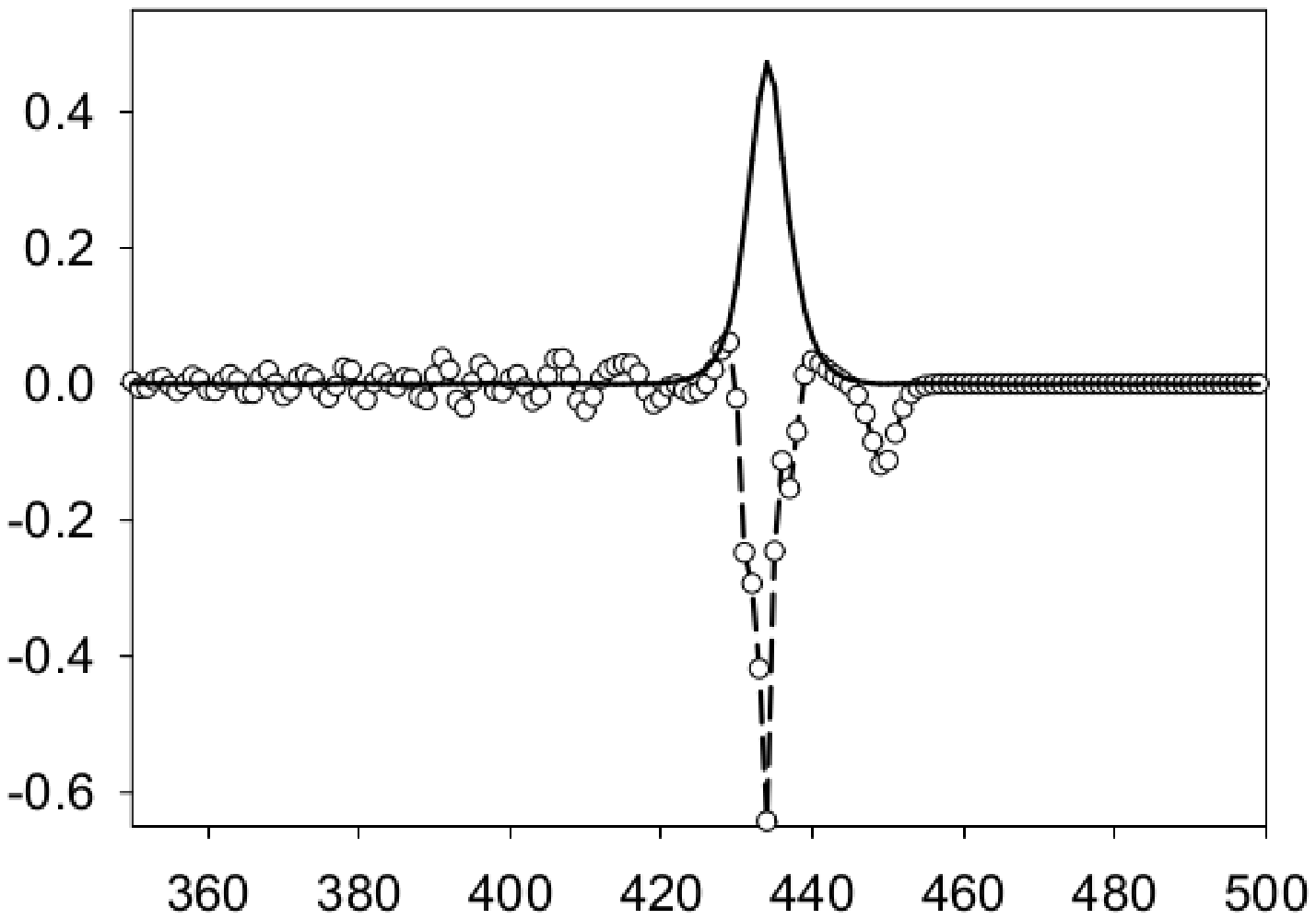}
  \includegraphics[width=80 mm]{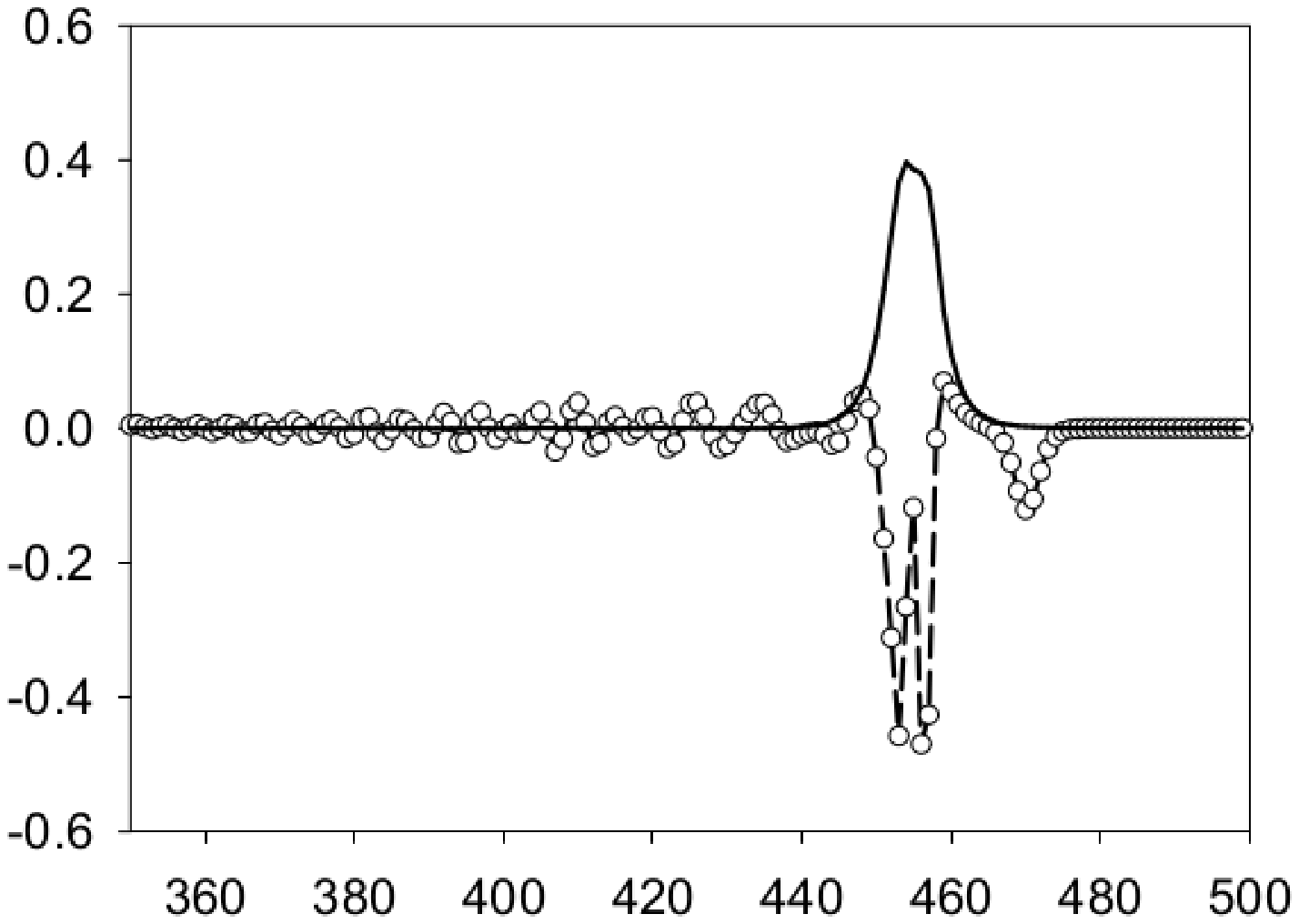}
 c) \hspace{8 cm} d)
  \caption{
The snapshots of the polaron at different time moments: $t = 321$
(a), $t= 345$ (b),$t = 380$ (c), $t = 400$ (d).
        }
   \label{Fig_09}
 \end{center}
\end{figure}

Fig.~\ref{Fig_09} shows how the two-peaked polaron self-organizes.
The initial excitation is chosen more narrow than in the previous
numerical experiment. The initial parameters are $d = 0.5$, $A =
0.6$, $v_{\rm p} = 0.7$, $k=0$. The evolution of this initial
state is more complex: there are intermediate one--, two-- and
three-peaked excitations. And finally, after rather long time, the
two-peaked polaron becomes the stable state. Its velocity is
somewhat less then the sound velocity. Numerical simulations show
that other initial conditions can produce three-peaked polarons.

And finally we consider the inharmonic lattice with parameters
$\alpha$ and $\beta$ corresponding to the parameters in DNA.
Usually the Morse potential is used to account the interaction of
neighboring bases $U(q) = D [1 - \exp(- b q)]^2$. An expansion of
this potential into the Taylor series gives the $\alpha$-FPU
potential. In the dimensionless form the potential is $U(q)
\approx \dfrac12 q^2 - \dfrac{\beta}{3} q^3$ with $\beta \approx
1.1$ and $\alpha \approx 1.2$.

\begin{figure}
 \begin{center}
  \includegraphics[width=53 mm]{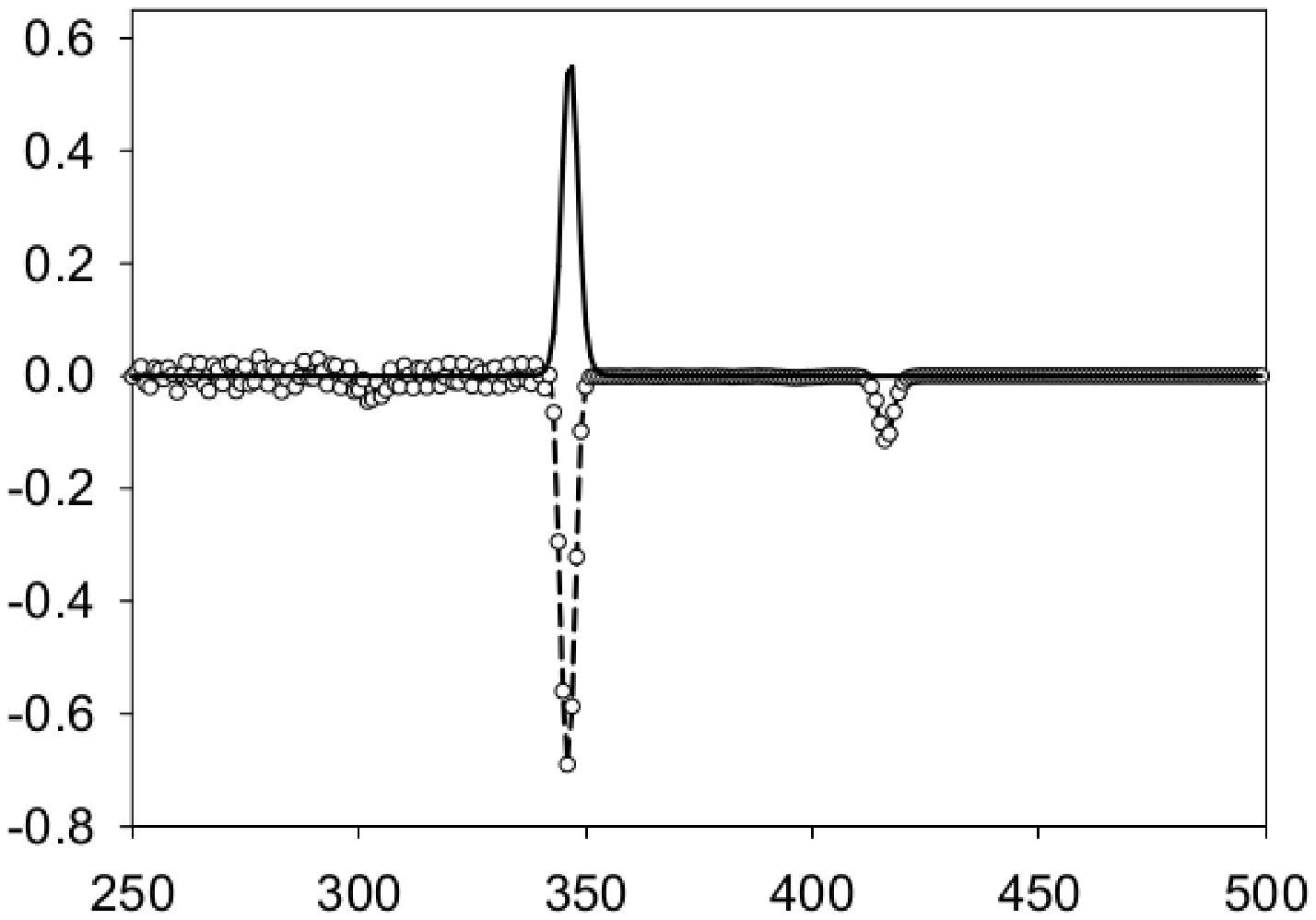}
  \includegraphics[width=53 mm]{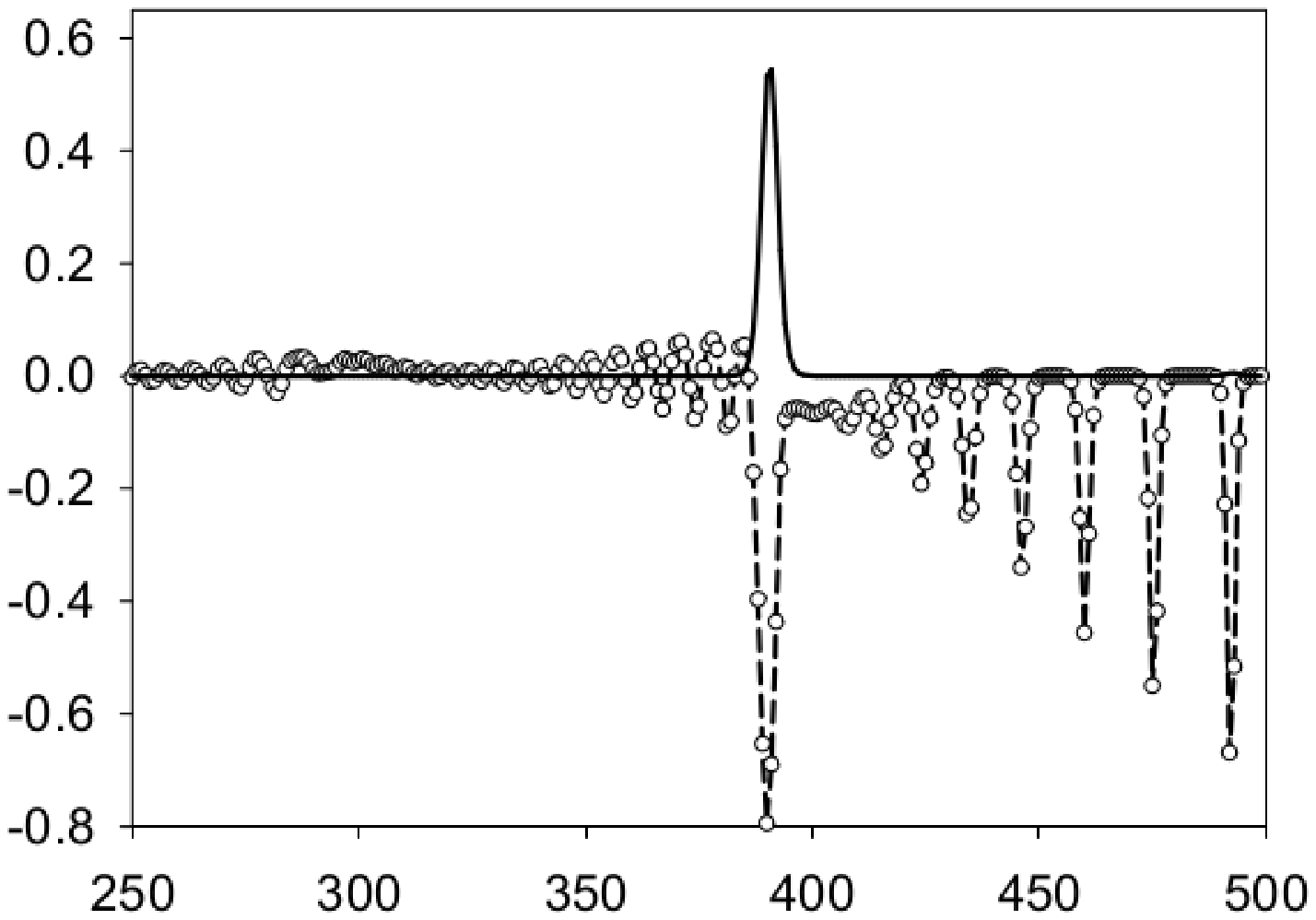}
  \includegraphics[width=53 mm]{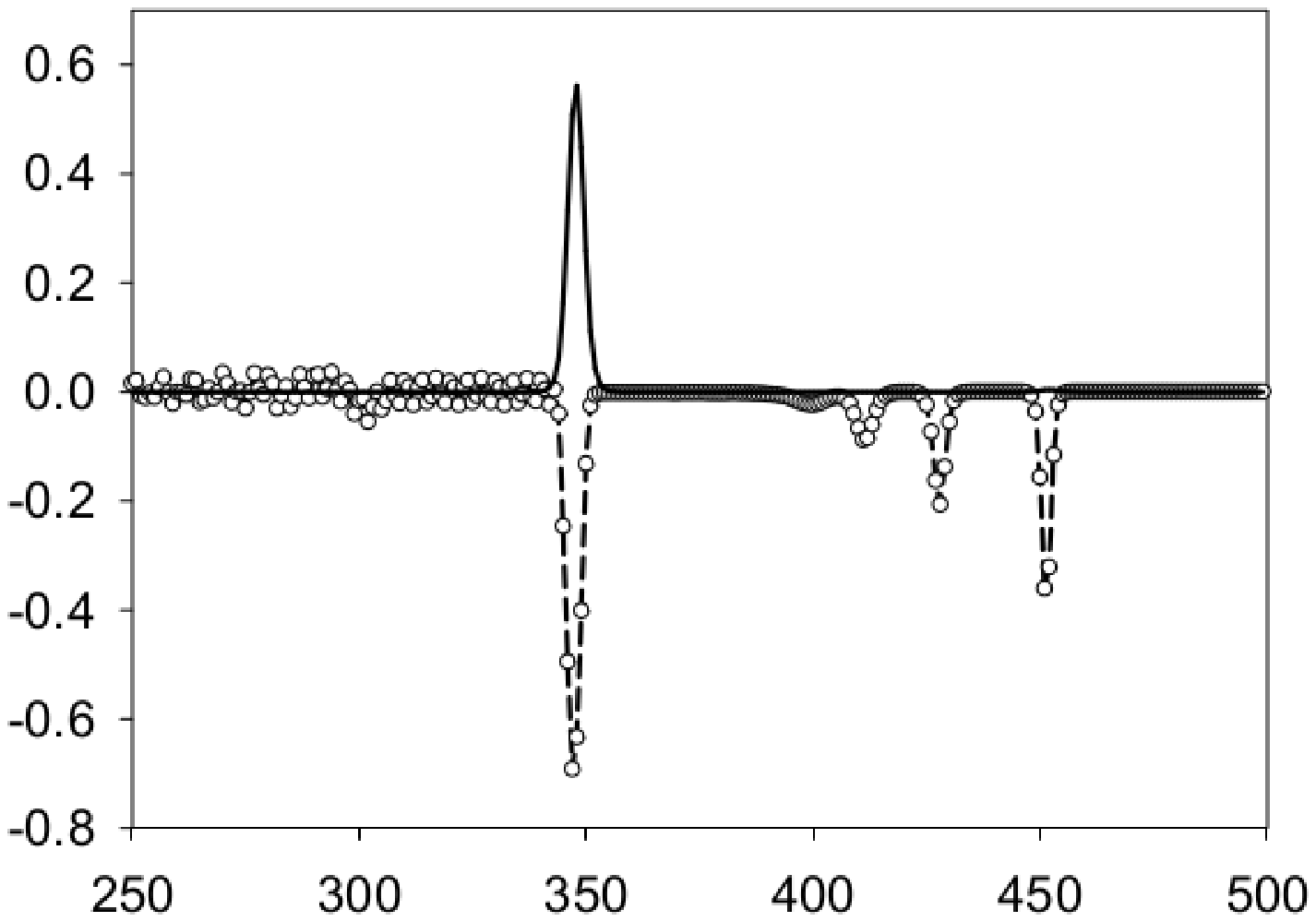}
 a) \hspace{4.5 cm} b)  \hspace{4.5 cm} c)
  \caption{
The result of evolution of the excitation with different initial
parameters at $t = 350$: $A = 0.6, \,\, d = 0.4, \,\, v_{\rm p}
=1.0$ (a); $A = 0.6, \,\, d = 0.1, \,\, v_{\rm p} =1.2$ (b); $A =
0.5, \,\, d = 0.2, \,\, v_{\rm p} =0.5$ (c). Lattice parameters:
$N = 500$, $\alpha = 1.2$, $\beta = 1.1$.
        }
   \label{Fig_10}
 \end{center}
\end{figure}

This set of parameters does not allow an analytical consideration.
Numerical simulations are carried out for wide range of initial
conditions. As an examples the evolution with different initial
parameters are analyzed. Results are shown in Fig.~\ref{Fig_10}.
Polarons self-organize from all used initial conditions. But no
multi-peaked polarons are found at these parameters choice.

The obtained results are rather unusual. First, the polaron can be
supersonic with velocity exceeding the fastest soliton velocity.
Second, the polaron envelope can be multi-peaked and consists of
two, three and four peaks (multi-peaked polarons were found
earlier in \cite{Fue04} in a slightly different problem
formulation). It means that the minimum energy states can have a
breather (multi-peaked) or a solitonic (single-peaked) character
and depend on the parameter values and initial conditions. Third,
the wave function is fully concentrated inside the self-organized
polaronic potential well, and emitted solitons do not carry away a
minute amount of a wave function.

Polarons on the homogeneous lattice were considered above. But DNA
and polypeptides can have static and dynamical defects of
different nature. An interaction of polarons with lattice defects
is considered in the next section.


\section{Conclusions}

In conclusion we briefly summarize the main results.

The moving polarons on the harmonic and inharmonic lattices are
considered. An analytical solutions for the moving polarons are
derived in the continuous approximation. This approximation is
valid for the large-radius polarons and when the electron-phonon
interaction $\alpha$ is weak and the nonlinearity parameter
$\beta$ is small. In few limiting cases these equations are
reduced to the exactly solvable models with special one-soliton
solutions. The polaron on the harmonic lattice is subsonic. The
polaron on the inharmonic lattice can travel with the supersonic
velocity. An excellent agreement between analytical results and
numerical simulation was obtained.

In the intermediate range of parameters $\alpha$ and $\beta$,
where the continuous approximation is not valid,  the family of
stable solutions with unexpected properties was found in numerical
simulations. These solutions are supersonic with velocities
exceeding the velocity of the fastest solitons. The envelop of
these polarons for the relative displacements consists of few
peaks. These polarons are similar to polarobreathers as they have
internal vibrational structure. For the larger parameter values,
typical for DNA, all formed polarons are subsonic with the
bell-shaped envelope.

In all cases the wave function is concentrated with the 100\%
probability inside the polaron potential well.

The results can be useful in explaining the recent experiments on
the highly efficient charge transfer in synthetic
oligonucleotides, where the CT occurs as the single-step coherent
process.


\end{document}